\documentclass[12pt]{article}
\usepackage[dvips]{graphicx}  
\begin{document}

\small{.}\vspace{15mm}

\begin{center}{\bf \Huge \hspace{5mm}ABSOLUTE MOTION } 
\vspace{8mm}

{\bf \Large AND }

\vspace{5mm}

{\bf \Huge  GRAVITATIONAL 

\vspace{6mm}

EFFECTS}

\vspace{20mm}
 
{{\bf \LARGE Reginald T. Cahill}}

\vspace{20mm}

      {\bf \Large School of Chemistry, Physics and Earth Sciences}
\vspace{7mm}

      {\bf \Large  Flinders University }
\vspace{7mm}

      {\bf \large GPO Box 2100, Adelaide 5001, Australia }
\vspace{15mm}

      {\bf \large Reg.Cahill@flinders.edu.au}
\vspace{5mm}

{\bf Process Physics URL: }
      
{\bf http://www.scieng.flinders.edu.au/cpes/people/cahill\_r/processphysics.html}

\vspace{5mm}

{\bf - June 2003 -}

\end{center}
\newpage

\vspace{10mm}
\begin{center}
{\hspace*{800mm} \\ \bf \large  Abstract}
\end{center}
\vspace{10mm}
{\small The new Process Physics  provides a new explanation of space as a quantum foam system in which gravity
is an inhomogeneous flow of the quantum foam into matter.  An analysis of various experiments demonstrates that
absolute motion  relative to space  has been observed experimentally by Michelson
and Morley, Miller, Illingworth,  Torr and Kolen, and by  DeWitte.  The  Dayton Miller and Roland DeWitte data also reveal the
in-flow of space into matter which manifests as  gravity.    The in-flow also  manifests turbulence and the experimental data
confirms this as well, which amounts to the observation of a gravitational wave phenomena. The Einstein assumptions
leading to the Special and General Theory of Relativity are shown to be falsified by the extensive experimental
data.  Contrary to the Einstein assumptions absolute motion is consistent with relativistic effects, which are
caused by actual dynamical effects of absolute motion through the quantum foam, so that it is Lorentzian 
relativity that is seen to be essentially  correct. }

\vspace{25mm}

  gravity, absolute motion

\newpage

 \tableofcontents
\newpage

\section{  Introduction\label{section:introduction}}

 The new Process Physics \cite{RCPP2003,RCGQF1,RC01,RC02,CK97,CK98,CK99,CKK00,CK,RC03,RC05,Kitto,MC}  provides a
new explanation of space as a quantum foam system in which gravity is an inhomogeneous flow of the quantum foam
into matter.  Here an analysis of data from various experiments demonstrates that
absolute motion  relative to space  has been observed  by Michelson
and Morley, Miller, Illingworth,  Torr and Kolen, and by  DeWitte, contrary to common belief within physics that
absolute motion has never been observed.   The  Dayton Miller and Roland DeWitte data also reveal the in-flow of space
into matter which manifests as  gravity.    The experimental data suggests
that the in-flow manifest turbulence, which amounts to the observation of a gravitational wave phenomena. The Einstein
assumptions leading to the Special and General Theory of Relativity are shown to be falsified by the extensive
experimental data.

Contrary to the Einstein assumptions absolute motion is consistent with relativistic effects, which are
caused by actual dynamical effects of absolute motion through the quantum foam.  Lorentzian 
relativity  is seen to be essentially  correct.

This paper is a condensed version of certain sections of Cahill \cite{RCPP2003}.

\section{ Detection of  Absolute Motion \label{section:detectionofabsolute}}

\subsection{Space and Absolute Motion\label{subsection:spaceandabsolute}}

Absolute motion is motion relative to space itself.  It turns out that Michelson and Morley in their historic
experiment of 1887 did detect absolute motion, but rejected their own findings because using Galilean relativity the
determined speed of some 8 km/s was less than the 30 km/s orbital speed of the Earth.  The data was clearly indicating
that the theory for the operation of the Michelson interferometer was not adequate.  Rather than reaching this
conclusion Michelson and Morley came to the incorrect conclusion that their results amounted to the failure to detect
absolute motion.   This had an enormous impact on the deveopment of physics, for as is well known Einstein adopted the
absence  of absolute motion effects as one of his fundamnetal assumptions.  By the time Miller had finally figured out
how to use and properly analyse data from his Michelosn interferometer absolute motion had become a forbidden concept
within physics, as it still is at present.  The  experimental observations  by Miller and others of absolute motion has
continued to be scorned and rejected by the physics community.  Unfortunately as well as revealing absolute motion the
experimental data also reveals evidence in support of a new theory of gravity.   

\subsection{  Theory of the  Michelson Interferometer\label{subsection:theoryof}}

\begin{figure}[h]

\setlength{\unitlength}{1.0mm}
\hspace{28mm}\begin{picture}(0,30)
\thicklines
\put(-10,0){\line(1,0){50}}
\put(-5,0){\vector(1,0){5}}
\put(40,-1){\line(-1,0){29.2}}
\put(15,0){\vector(1,0){5}}
\put(30,-1){\vector(-1,0){5}}
\put(10,0){\line(0,1){30}}
\put(10,5){\vector(0,1){5}}
\put(11,25){\vector(0,-1){5}}
\put(11,30){\line(0,-1){38}}
\put(11,-2){\vector(0,-1){5}}
\put(8.0,-2){\line(1,1){5}}
\put(9.0,-2.9){\line(1,1){5}}
\put(6.5,30){\line(1,0){8}}
\put(40,-4.5){\line(0,1){8}}
\put(5,12){ $L$}
\put(4,-5){ $A$}
\put(35,-5){ $B$}
\put(25,-5){ $L$}
\put(12,26){ $C$}
\put(9,-8){\line(1,0){5}}
\put(9,-9){\line(1,0){5}}
\put(14,-9){\line(0,1){1}}
\put(9,-9){\line(0,1){1}}
\put(15,-9){ $D$}
\put(50,0){\line(1,0){50}}
\put(55,0){\vector(1,0){5}}
\put(73,0){\vector(1,0){5}}
\put(85,0){\vector(1,0){5}}
\put(90,15){\vector(1,0){5}}
\put(100,-1){\vector(-1,0){5}}
\put(100,-4.5){\line(0,1){8}}
\put(68.5,-1.5){\line(1,1){4}}
\put(69.3,-2.0){\line(1,1){4}}
\put(70,0){\line(1,4){7.5}}
\put(70,0){\vector(1,4){3.5}}
\put(77.5,30){\line(1,-4){7.7}}
\put(77.5,30){\vector(1,-4){5}}
\put(73.5,30){\line(1,0){8}}
\put(83.3,-1.5){\line(1,1){4}}
\put(84.0,-2.0){\line(1,1){4}}
\put(100,-1){\line(-1,0){14.9}}
\put(67,-5){ $A_1$}
\put(82,-5){ $A_2$}
\put(95,-5){ $B$}
\put(79,26){ $C$}
\put(90,16){ $v$}
\put(-8,8){(a)}
\put(55,8){(b)}

\end{picture}

\vspace{10mm}
\caption{\small{Schematic diagrams of the
Michelson Interferometer, with
beamsplitter/mirror at $A$ and mirrors at $B$ and
$C$, on equal length arms when parallel, from $A$. $D$ is a quantum detector (not drawn in (b)) that causes
localisation  of the photon state by a collapse process. In (a) the interferometer is at
rest in space. In (b) the interferometer is moving with speed $v$ relative to space in
the direction indicated. Interference fringes are observed at the quantum detector $D$. 
If the interferometer is rotated in the plane  through $90^o$, the roles of arms
$AC$ and $AB$ are interchanged, and during the rotation shifts of the fringes are seen
in the case of absolute motion, but only if the apparatus operates in a gas.  By counting
fringe changes the speed $v$ may be determined.}\label{fig:Minterferometer}}
\end{figure}
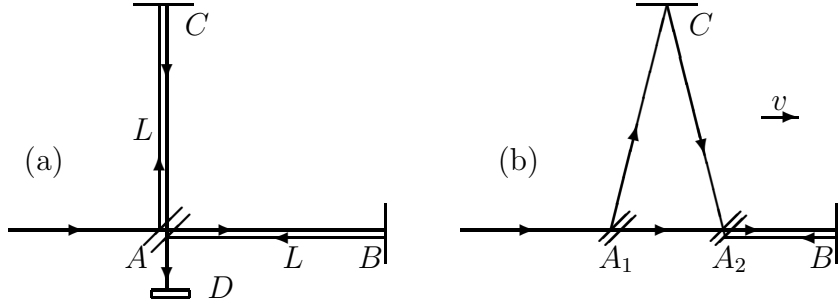

We now show for the first time in over 100 years how the  three key effects together permit the 
Michelson interferometer \cite{Mich} to  reveal the phenomenon of absolute
motion when operating in the presence of a gas, with the third effect only discovered in 2002
\cite{CK}.       The main outcome is the  derivation of the origin of the Miller
$k^2$  factor in the expression for the time difference for light travelling via the orthogonal arms,
\begin{equation}\label{eqn:QG0}
\Delta t=2k^2\frac{L|{\bf v}_P|^2}{c^3}\cos(2(\theta-\psi)).
\end{equation}
 Here ${\bf v}_P$ is the projection of the absolute velocity ${\bf v}$ of the
interferometer  through the quantum-foam  onto the plane of the interferometer, where the projected
velocity vector ${\bf v}_P$ has azimuth angle $\psi$ relative to the local meridian, and
$\theta$ is the angle of one arm from that meridian.  The  $k^2$ factor is   $k^2=n(n^2-1)$  where $n$
is the refractive index of the gas through which the light  passes,  $L$ is the
length of each arm and
$c$ is the speed of light relative to the quantum foam. This expression follows from three key effects: (i)
the difference in geometrical length of the two paths when the interferometer is in absolute motion, as
first realised by Michelson,  (ii) the Fitzgerald-Lorentz contraction of the arms along the direction of
motion, and (iii) that these two effects precisely cancel in vacuum, but leave a residual effect if
operated in a gas,  because the speed of light through the gas is reduced to
$V=c/n$, ignoring here for simplicity any  Fresnel-drag effects, see \cite{RCPP2003}. 
This is one of the  aspects of the
quantum foam physics that distinguishes it from the Einstein formalism.  The time difference
$\Delta t$ is revealed by the fringe shifts on rotating the interferometer. In Newtonian physics,
that is with no Fitzgerald-Lorentz contraction, $k^2=n^3$, while in Einsteinian physics $k=0$
reflecting the fundamental assumption that absolute motion is not measurable and indeed has no
meaning. The Special Relativity null effect for the interferometer is explicitly derived in
\cite{RCPP2003}.  So the experimentally determined value of
$k$ is a key test of fundamental physics.  Table 1 summarises the differences between the three fundamental
theories in their  modelling of time, space, gravity and the quantum, together with their 
distinctive values for the interferometer parameter $k^2$.  For  air $n=1.00029$, and so for
process physics $k=0.0241$ and $k^2=0.00058$, which is close to the Einsteinian value of $k=0$,
particularly in comparison to the Newtonian value of $k=1.0$.  This small but non-zero $k$ value explains
why the Michelson interferometer experiments gave such small fringe shifts.  Fortunately it is possible to
check  the $n$ dependence of $k$ as one experiment \cite{Illingworth} was done in Helium gas, and this has
an $n^2-1$ value significantly different from that of air. 

\vspace{3mm}
{\footnotesize
\hspace{-3mm}\begin{tabular}{| l|c |c|c|c|c|} 
\hline 
{ \bf Theory} &  Time & Space  & Gravity & Quantum &$k^2$ \\
\hline\hline  
{\bf Newton}   & geometry  & geometry & force & Quantum Theory & $n^3$ \\ \hline
 
{\bf Einstein}  & \multicolumn{2}{c}{curved geometry} \vline & curvature & Quantum Field Theory &0
\\ \hline 
{\bf Process}   & process & quantum & inhomogeneous  & Quantum Homotopic  &$n(n^2-1)$
\\
 & &foam & flow & Field Theory & \\ 
\hline
\end{tabular}}

\vspace{2mm}
Table 1: {\small Comparisons of Newtonian, Einsteinian and Process Physics. } 

\vspace{2mm}
  
In deriving  (\ref{eqn:QG1}) in the new physics it is essential to note that space is a quantum-foam system
  which exhibits various subtle features. In particular it exhibits real dynamical
effects on clocks and rods. In this physics the speed of light is only $c$  relative to the
quantum-foam, but to observers moving with respect to this quantum-foam the speed appears to be still $c$, but
only because their clocks and rods are affected by the quantum-foam. As shown in 
\cite{RCPP2003} such observers will find that records of observations of
distant events will be described by the Einstein spacetime formalism, but only if they restrict
measurements to those achieved by using clocks, rods and light pulses, that is using the Einstein
measurement protocol.   However if they use an absolute motion detector then such observers can correct
for these effects. 

It is simplest
in the new physics to work in the quantum-foam frame of reference.  If there is a gas present at rest in
this frame, such as air, then the speed of light in this frame is
$V=c/n$. If the interferometer and gas  are moving with respect to the quantum foam, as in the case of an
interferometer attached to the Earth, then the speed of light relative to the quantum-foam is still
$V=c/n$ up to corrections due to  drag effects.    Hence this new physics requires a different method of analysis
from that of the Einstein physics. With these cautions we now describe the operation of a Michelson
interferometer in this new physics, and show that it makes predictions different to that of the Einstein
physics.    Of course experimental evidence is the final arbiter in this conflict of theories.

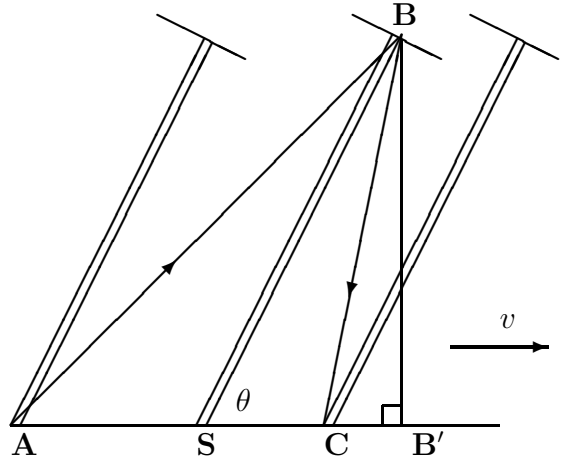
\begin{figure}[ht]
\vspace{15mm}\setlength{\unitlength}{1.3mm}
\hspace{35mm}\begin{picture}(20,30)
\thicklines
\put(0,0){\line(1,0){50}}
\put(0,-3){{\bf A}}
\put(32,0){\line(1,5){8}}

\put(32,-3){{\bf C}}
\put(19,0){\line(1,2){20.0}}
\put(20,0){\line(1,2){20}}
\put(19,-3){{\bf S}}
\put(23,+1.5){{$\theta$}}

\put(35,42){\line(2,-1){9}}
\put(39,41){{\bf B}}
\put(0,0){\line(1,1){39.6}}
\put(50,10){{$ v$}}
\put(45,8){\vector(1,0){10}}

\put(0,0){\line(1,2){19.8}}
\put(1,0){\line(1,2){19.6}}
\put(15,42){\line(2,-1){9}}

\put(32,0){\line(1,2){19.8}}
\put(33,0){\line(1,2){19.6}}
\put(47,42){\line(2,-1){9}}

\put(40,0){\line(0,1){39.5}}
\put(41,-3){{\bf B}$'$}

\put(38,0){\line(0,1){2}}
\put(38,2){\line(1,0){2}}

\put(15,15){\vector(1,1){2}}
\put(35.02,15){\vector(-1,-4){0.5}}

\end{picture}

\vspace{3mm}
\caption{\small{One arm  of a Michelson Interferometer travelling  at  angle $\theta$ and   velocity 
${\bf v}$, and shown at three successive times: (i) when photon leaves beamsplitter at $A$, (ii) when photon
is reflected at mirror $B$, and (iii) when photon returns to beamsplitter at $C$. The line $BB'$ defines
right angle triangles $ABB'$ and $SBB'$.  The second arm is not shown but has angle $\theta+90^o$ to 
${\bf v}$. Here ${\bf v}$ is in the plane of the interferometer for simplicity, and
the azimuth angle $\psi=0$. }\label{fig:MMangled}}
\end{figure}

As shown in Fig.\ref{fig:MMangled}  the  beamsplitter/mirror when  at $A$ sends a photon $\psi(t)$ into a
superposition
$\psi(t)=\psi_1(t)+\psi_2(t)$, with each component travelling in different arms of the interferometer, until
they are recombined in the quantum detector which results in a localisation process, and one spot in the
detector is produced.  Repeating with many photons reveals that the
interference between
$\psi_1$ and
$\psi_2$ at the detector results in fringes.  These fringes actually only appear if the mirrors are not quite
orthogonal, otherwise the screen has  a uniform intensity and this intensity changes as the interferometer
is rotated, as shown in the analysis by  Hicks \cite{Hicks}.   To simplify the analysis here assume that
the two arms are constructed to have the same lengths
$L$  when they are physically parallel to each other and perpendicular to $ v$, so that the distance 
$BB'$ is
$L\sin(\theta)$. The Fitzgerald-Lorentz effect in the new physics  is that the distance  $SB'$  is 
$\gamma^{-1} L\cos(\theta)$ where
$\gamma=1/\sqrt{1-v^2/c^2}$.  The various other distances  are $AB=Vt_{AB}$, $BC=Vt_{BC}$, $AS=vt_{AB}$ 
and $SC=vt_{BC}$, where $t_{AB}$ and $t_{BC}$ are the travel times.  Applying the Pythagoras theorem to
triangle $ABB'$ we obtain
\begin{equation}\label{eqn:QG1}
t_{AB}=\frac{2v\gamma^{-1}
L\cos(\theta)+\sqrt{4v^2\gamma^{-2}L^2\cos^2(\theta)+4L^2(1-\frac{v^2}{c^2}\cos^2(\theta))(V^2-v^2)}}{2(V^2-v^2)}.
\end{equation}

\noindent The expression for $t_{BC}$ is the same except for a change of sign of the  $2v\gamma^{-1}
L\cos(\theta)$ term, then  
\begin{equation}\label{eqn:QG2}  
t_{ABC}=t_{AB}+t_{BC}=\frac{\sqrt{4v^2\gamma^{-2}L^2\cos^2(\theta)+4L^2(1-\frac{v^2}{c^2}\cos^2(\theta))
(V^2-v^2)}}{(V^2-v^2)}.
\end{equation}
The  corresponding travel time $t'_{ABC}$ for the orthogonal arm  is obtained from  (\ref{eqn:QG2}) by the
substitution $\cos(\theta) \rightarrow \cos(\theta+90^0)=\sin(\theta)$. The difference in travel times
between the two arms is then $\Delta t= t_{ABC}-t'_{ABC}$. Now trivially $\Delta t =0$  if $v=0$, but 
also $\Delta t =0$ when
$v\neq 0$ but only if $V=c$.  This then would  result in a null result on rotating the apparatus.  Hence the null
result of  Michelson interferometer  experiments in the new physics is only for the special case of
photons travelling in vacuum for which $V=c$.    However if the interferometer is immersed
 in a gas then $V<c$ and a non-null effect is expected on rotating the apparatus, since now 
$\Delta t \neq 0$.  It is essential then in analysing data to correct for this refractive index effect.  The above
$\Delta t$ is the change in travel time when one arm is moved through angle $\theta$.  The interferometer
operates by comparing the change in the difference of the travel times between the arms, and this introduces a
factor of 2. Then for
$V=c/n$ we  find for $v << V$  that 
\begin{equation}\label{eqn:QG4}
\Delta t= 2Ln(n^2-1)\frac{v^2}{c^3}\cos(2\theta)+\mbox{O}(v^4),
\end{equation}
that is $k^2=n(n^2-1)$,  which gives $k=0$ for vacuum experiments ($n=1$).  So the Miller
phenomenological parameter $k$  is seen to accommodate  both the Fitzgerald-Lorentz
contraction effect and the dielectric effect, at least for gases\footnote{For
Michelson interferometers using light propagation through solids such as plastic or optical
fibres there is evidence, discussed in \cite{RCPP2003}, that  another effect comes into
operation, namely a non-isotropic change of refractive index that causes absolute motion effects to be
completely cancelled.}.  This is very fortunate since being a multiplicative parameter a re-scaling of old
analyses is all that is required. $\Delta t$ is non-zero  when
$n \neq 1$ because the refractive index effect results in incomplete cancellation of the geometrical
effect and the Fitzgerald-Lorentz contraction effect.  Of course it was this cancellation effect that
Fitzgerald and Lorentz actually used to arrive at the length  contraction hypothesis, but they failed to
take the next step and note that the cancellation would be incomplete in the air operated Michelson-Morley
experiment.  In a bizarre development modern Michelson interferometer experiments, which use resonant
cavities rather than interference effects, but for which the analysis here is easily adapted, and with the
same consequences, are operated in vacuum mode.  That denies these experiments the opportunity to see
absolute motion effects.  Nevertheless the experimentalists continue to misinterpret their null results as
evidence against absolute motion.  Of course  these experiments are therefore restricted to merely checking the
Fitzgerald-Lorentz contraction effect, and this is itself of some interest.

All data from  gas-mode interferometer experiments, except for that of Miller,  has been incorrectly 
analysed using only the first effect as in Michelson's initial theoretical treatment, and so the
consequences of the other two effects have been absent.  Repeating the above analysis without these two
effects  we arrive at the  Newtonian-physics time difference which,  for $v << V$, is 
\begin{equation}\label{eqn:QG5}
\Delta t =2 Ln^3\frac{v^2}{c^3}\cos(2\theta)+\mbox{O}(v^4),
\end{equation}
that is $k^2=n^3$. The value of $\Delta t$, which  is typically of order $10^{-17}s$  in gas-mode
interferometers corresponding to a fractional fringe
shift, is deduced from analysing the fringe shifts, and then    the speed $v_{M}$  has been
extracted  using (\ref{eqn:QG5}), instead of the correct form (\ref{eqn:QG4}) or more generally
(\ref{eqn:QG1}).    However it is very easy to correct for this oversight.  From (\ref{eqn:QG4}) and
(\ref{eqn:QG5}) we obtain for the corrected absolute (projected) speed 
$v_P$ through space, and for $n \approx 1^+$, 
\begin{equation}\label{eqn:QG6}
v_P=\frac{v_{M}}{\sqrt{n^2-1}}.
\end{equation}
For air the correction factor in (\ref{eqn:QG6}) is significant, and even more so for Helium.

\subsection{  The Michelson-Morley Experiment:  1887\label{subsection:themichelsonmorley}}

 Michelson and Morley reported  that their interferometer experiment in 1887  gave a
`null-result' which  since then, with rare exceptions, has been claimed to  support the Einstein
assumption that absolute motion has no meaning.  However to the contrary  the Michelson-Morley published
data \cite{MM} shows non-null effects, but much smaller than they expected.  They made observations of
thirty-six  $360^0$ turns  using an $L=11$ meter length interferometer  operating in air in Cleveland
(Latitude $41^0  30'$N) with six turns near 
$12\!:\!\!00$ hrs ($7\!\!:\!\!00$ hrs ST) on each day of July 8, 9 and 11, 1887  and similarly near
$18\!:\!\!00$ hrs ($13\!\!:\!\!00$ hrs ST) on July 8, 9 and 12, 1887.  Each turn took approximately 6
minutes as the interferometer slowly rotated floating on a tank of mercury.  They published and analysed
the average of each of the  6 data sets.  The fringe shifts were extremely small but within their
observational capabilities.

\begin{figure}
\hspace{15mm}\includegraphics[scale=1.5]{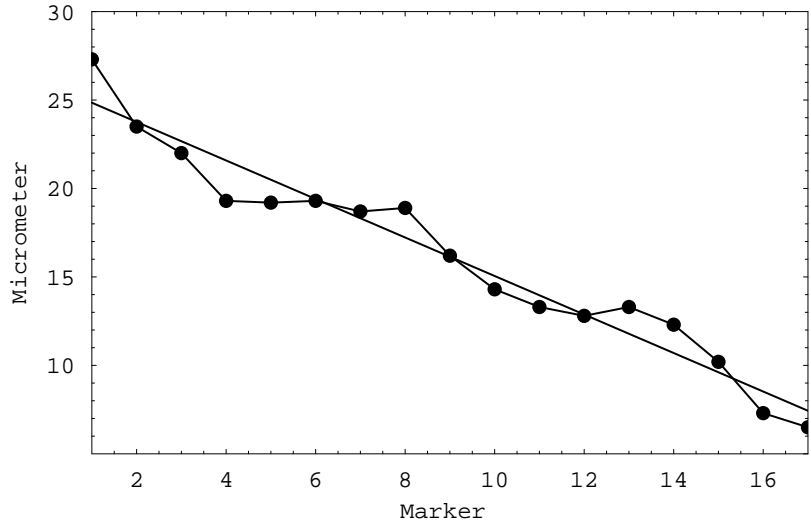}
\caption{{\small   Plot of micrometer  readings for July 11 $12\!\!:\!\!00$ hr ($7\!\!:\!\!00$ ST) showing the
absolute motion induced fringe shifts superimposed on the uniform temperature induced
fringe drift.}\label{fig:MMrawdata}}
\end{figure}

\begin{figure} 
\hspace{10mm}{\small \begin{tabular}{|c|c|c|c|c|c|c|c|c|c|}
\hline
 local &16& 1& 2& 3& 4& 5& 6& 7& \\time & 8  &  9  & 10 &11 &12 & 13 & 14 & 15 & 16 \\
\hline
 12:00hr &27.3 & 23.5 & 22.0 & 19.3& 19.2& 19.3& 18.7& 18.9 &\\July 11 & 16.2 
&14.3& 13.3& 12.8& 13.3& 12.3& 10.2& 7.3& 6.5  \\ 
\hline
 18:00hr &26.0& 26.0& 28.2& 29.2& 31.5& 32.0& 31.3& 31.7& \\July 9 & 33.0
& 35.8& 36.5& 37.3& 38.8&
41.0& 42.7& 43.7& 44.0 \\
\hline
 \end{tabular}}
\vspace{3mm}

\hspace{6mm}{\small Table 2. Examples of Michelson-Morley fringe-shift micrometer readings. 

\hspace{5mm} The readings for July 11  $12\!\!:\!\!00$ hr are plotted in Fig.\ref{fig:MMrawdata}.}
\end{figure}

\vspace{3mm}

Table 2 shows   examples of the averaged fringe shift micrometer readings every $22.5^0$ of rotation of
the   Michelson-Morley interferometer \cite{MM} for July 11 12:00 hr  local time and also for July 9
18:00 hr local time.  The orientation of the  stone slab  base is
 indicated by the marks $16,1,2,..$. North is mark 16. The dominant effect was a uniform
fringe drift caused by temporal temperature effects on the length of the arms, and imposed upon that are
the fringe shifts corresponding to the effects of absolute motion, as shown in Fig.\ref{fig:MMrawdata}.

\begin{figure}
\hspace{25mm}\includegraphics[scale=1.25]{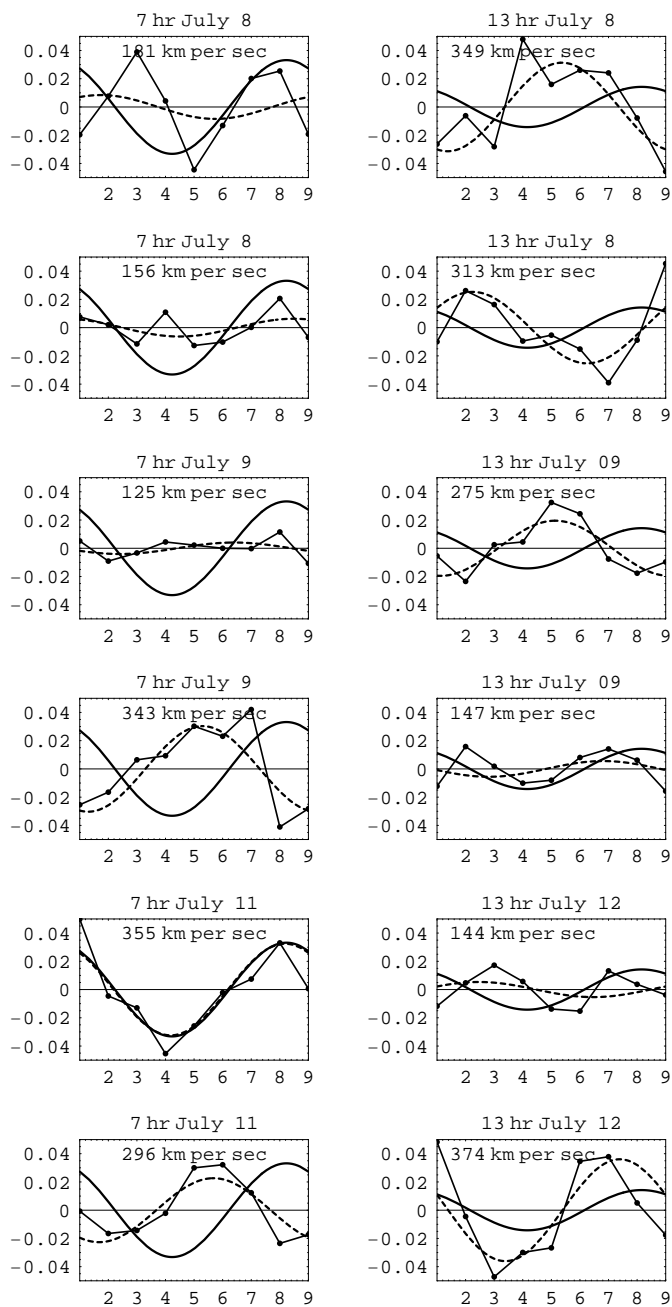}
\caption{{\small Shows all the Michelson-Morley 1887 data after removal of the
temperature induced fringe drifts.  The data for each $360^0$ full turn (the average of 6 individual turns) is
divided into the 1st and 2nd $180^0$ parts and  plotted one above the other.  The dotted curve
shows a best fit to the data, while the full curves show the expected forms using the
Miller direction for ${\bf v}_{cosmic}$.}\label{fig:MMplots}}
\end{figure}

This
temperature effect can be removed by subtracting from the data in each case a best fit to the data of
$a+bk$,
$\{k=0,1,2,.. ,8\}$ for the first $0^0$ to $180^0$  part of each rotation data set.   Then multiplying by
$0.02$ for the micrometer thread calibration   gives the fringe-shift
data points  in Fig.\ref{fig:MMplots}. This factor of $0.02$ converts the micrometer readings to fringe
shifts expressed as fractions of a wavelength.  Similarly a linear fit has been made to the data from the
$180^0$ to $360^0$  part of each rotation data set.  Separating  the full $360^0$ rotation into two $180^0$
parts  reduces the effect of the temperature drift not being perfectly linear in time.

\begin{figure}[ht]
\vspace{20mm}
\hspace{50mm}\setlength{\unitlength}{1.0mm}
\begin{picture}(0,20)
\thicklines
\put(25,10){\vector(-1,0){15}}
\put(25,10){\vector(0,-1){29.5}}
\put(25,10){\vector(-1,2){15}}
\qbezier(0,0)(25,20)(50,0)
\put(30,-15){\Large $\bf v_{in}$}
\put(10,-20){\Large $\bf Sun$}
\put(-10,10){\Large ${\bf v}_{tangent}$}
\put(17,30){\Large ${\bf v}_{R}$}
\end{picture}
\vspace{20mm} 
\caption{\small  Orbit of Earth about the Sun defining the  plane of the ecliptic with tangential orbital
velocity
${\bf v}_{tangent}$ and quantum-foam in-flow velocity  ${\bf v}_{in}$. Then ${\bf v}_{R}={\bf
v}_{tangent}-{\bf v}_{in}$ is the velocity of the Earth relative to the quantum foam, after subtracting
${\bf v}_{cosmic}$. }
\label{fig:orbit}
\end{figure}
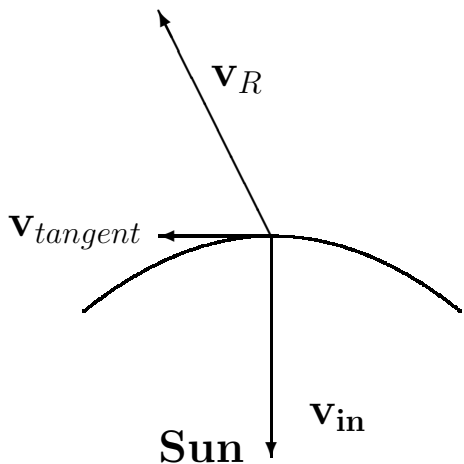

In the new physics there are four main velocities that contribute to the total velocity:
\begin{equation}\label{eqn:QG6b}
{\bf v}= {\bf v}_{cosmic} +{\bf v }_{tangent} -{\bf v}_{in}-{\bf v}_E.
\end{equation}
Here ${\bf v}_{cosmic}$ is the velocity of the Solar system relative to some cosmologically defined
galactic quantum-foam system (discussed later) while the other three are local effects: (i) ${\bf v
}_{tangent}$ is the tangential orbital velocity of the Earth about the Sun,  
(ii) ${\bf v}_{in}$ is  a quantum-gravity radial in-flow  of the quantum foam past the Earth towards the
Sun, and (iii) the corresponding quantum-foam in-flow into the
Earth is ${\bf v}_E$ and makes no contribution to a horizontally operated   interferometer, assuming the
velocity superposition approximation\footnote{and also that the
turbulence associated with that flow is not significant.} discussed in \cite{RCPP2003}. 
The minus signs in (\ref{eqn:QG6b}) arise because, for example, the in-flow towards the Sun requires the
Earth to have an outward directed velocity against that in-flow in order to maintain a fixed distance from
the Sun, as shown in Fig.\ref{fig:orbit}.   For circular orbits  and using in-flow form of Newtonian
gravity the speeds
$v_{tangent}$  and  $v_{in}$ are given by  
\begin{equation}\label{eqn:QG7}
v_{tangent}=\sqrt{\displaystyle{\frac{GM}{R}}},\end{equation}  
\begin{equation}\label{eqn:QG8}
v_{in}=\sqrt{\displaystyle{\frac{2GM}{R}}},\end{equation}
while the net speed $v_R$ of the Earth from the vector sum   ${\bf v}_R={\bf v}_{tangent}-{\bf v}_{in}$  is 
\begin{equation}\label{eqn:QG7b}
v_{R}=\sqrt{\displaystyle{\frac{3GM}{R}}},\end{equation}     
where $M$ is the mass of the Sun, $R$ is the distance of the Earth from the
Sun, and $G$ is Newton's gravitational constant. $G$ is essentially a measure of the rate at
which matter effectively `dissipates' the quantum-foam. The gravitational acceleration arises from 
inhomogeneities in the flow.  
 These expressions give $v_{tangent}=30$km/s,  $v_{in}=42.4$km/s and $v_{R}=52$km/s.

\begin{figure}
\hspace{25mm}\includegraphics[scale=1.2]{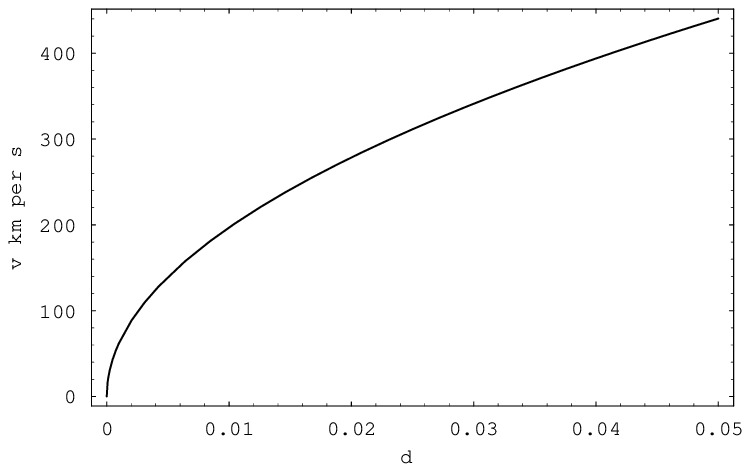}
\caption{{\small Speed calibration for Michelson-Morley experiment. This shows the value of $v_P$ in
km/s for values of the fringe shifts, $d$, expressed as a fraction of one wavelength of
the light used,  as shown in Fig.\ref{fig:MMplots}}\label{fig:MMcalib}}
\end{figure}

 Fig.\ref{fig:MMplots} shows all the data  for the 1887 Michelson-Morley experiment  for the fringe shifts after
removal of the temperature drift effect for each averaged  180 degree rotation. The dotted  curves  come from  
the best fit of $\frac{0.4}{30^2}k^2_{air}v_P^2\cos(2(\theta-\psi))$ to the data. The coefficient
$0.4/30^2$ arises as the apparatus would give a $0.4$ fringe shift, as a fraction of a wavelength,  with
$k=1$ if $v_P=30$ km/s \cite{MM}.  Shown in each figure is the resulting value of $v_P$.  In some cases the
data does not have the expected $\cos(2(\theta-\psi))$ form, and so the corresponding values for $v_P$ are
not meaningful.  The remaining fits give $v_P=331\pm30$ km/s for the   $7\!\!:\!\!00$ hr (ST)
data, and 
$v_P=328\pm50$ km/s for the $13\!\!:\!\!00$ hr (ST) data.   For comparison  the full curves show the
predicted form for the Michelson-Morley data, computed for the latitude of Cleveland, using the Miller
direction (see later) for  ${\bf v}_{cosmic}$ of Right Ascension and Declination $(\alpha=4^{hr} 54^\prime,
\delta=-70^0 30^\prime)$ and incorporating the tangential and in-flow velocity effects for July.   The
magnitude  of the theoretical curves are in general in good agreement with the magnitudes of the
experimental data, excluding those cases where the data does not have the sinusoidal form.  However there are significant
fluctuations in the  azimuth angle.  These fluctuations are also present in the Miller data, and together suggest that
this is a real physical phenomenon, and not solely due to difficulties with the operation of the interferometer.

The Michelson-Morley interferometer data clearly shows the characteristic sinusoidal form with period $180^0$ 
together with a large speed.  Ignoring the effect of the refractive index, namely using the Newtonian value of
$k=1$, gives speeds reduced by the factor $k_{air}$, namely  $k_{air}v_P=0.0241\times 330 $km/s $=7.9$ km/s.
Michelson and Morley reported  speeds in the range $5$km/s - $7.5$km/s. These slightly smaller speeds arise
because they averaged all the  $7\!\!:\!\!00$ hr (ST) data, and separately all the $13\!\!:\!\!00$ hr (ST) data,
whereas here some of the lower quality data has not been used.   Michelson was led to the false conclusion that
because this speed of some $8$ km/s was considerably less than the orbital speed of $30$ km/s  the
interferometer must have failed to have detected absolute motion, and that the data was merely caused by
experimental imperfections.  This was the flawed analysis that led to the incorrect conclusion by Michelson and
Morley that the experiment had failed to detect absolute motion.  The consequences for physics  were extremely
damaging, and are only now being rectified after some 115 years.

\subsection{  The Miller Interferometer Experiment: 1925-1926\label{subsection:themiller}}

Dayton Miller developed and operated  a  Michelson interferometer for over twenty years with an effective 
arm length of $L=32$m achieved by multiple
reflections.  The steel arms weighed 1200 kilograms and floated in a tank of
275 kilograms of Mercury. The main sequence of
observations being on Mt.Wilson in the years 1925-1926, with the results reported in 1933 by Miller \cite{Miller2}.  
  Miller developed his huge interferometer over the years, from 1902 to 1906 in
collaboration with Morley, and later at Mt.Wilson where the most extensive interferometer observations
were carried out.  Miller was meticulous in perfecting the operation of the interferometer and performed
many control experiments. The biggest problem to be controlled was the effect of temperature changes on
the lengths of the arms. It was essential that the temperature
effects were kept as small as possible, but so long as each turn was performed sufficiently quickly, any
temperature effect could be assumed to have been linear with respect to the angle of rotation. Then a
uniform background fringe drift could be removed,  as in the Michelson-Morley data analysis (see
Fig.\ref{fig:MMrawdata}).  

 In all some 200,000 readings were taken during some 12,000 turns of the
interferometer.  Analysis of the data requires the
extraction of the  speed
$v_M$ and the azimuth angle $\psi$ by effectively fitting the observed time differences, obtained from the
observed fringe shifts,  using (\ref{eqn:QG1}), but with $k=1$.  Miller was of course unaware of the full
theory of the interferometer and so he assumed the Newtonian theory, which neglected both  the
Fitzgerald-Lorentz contraction  and air effects.

\begin{figure}
\hspace{7mm}\includegraphics[scale=0.8]{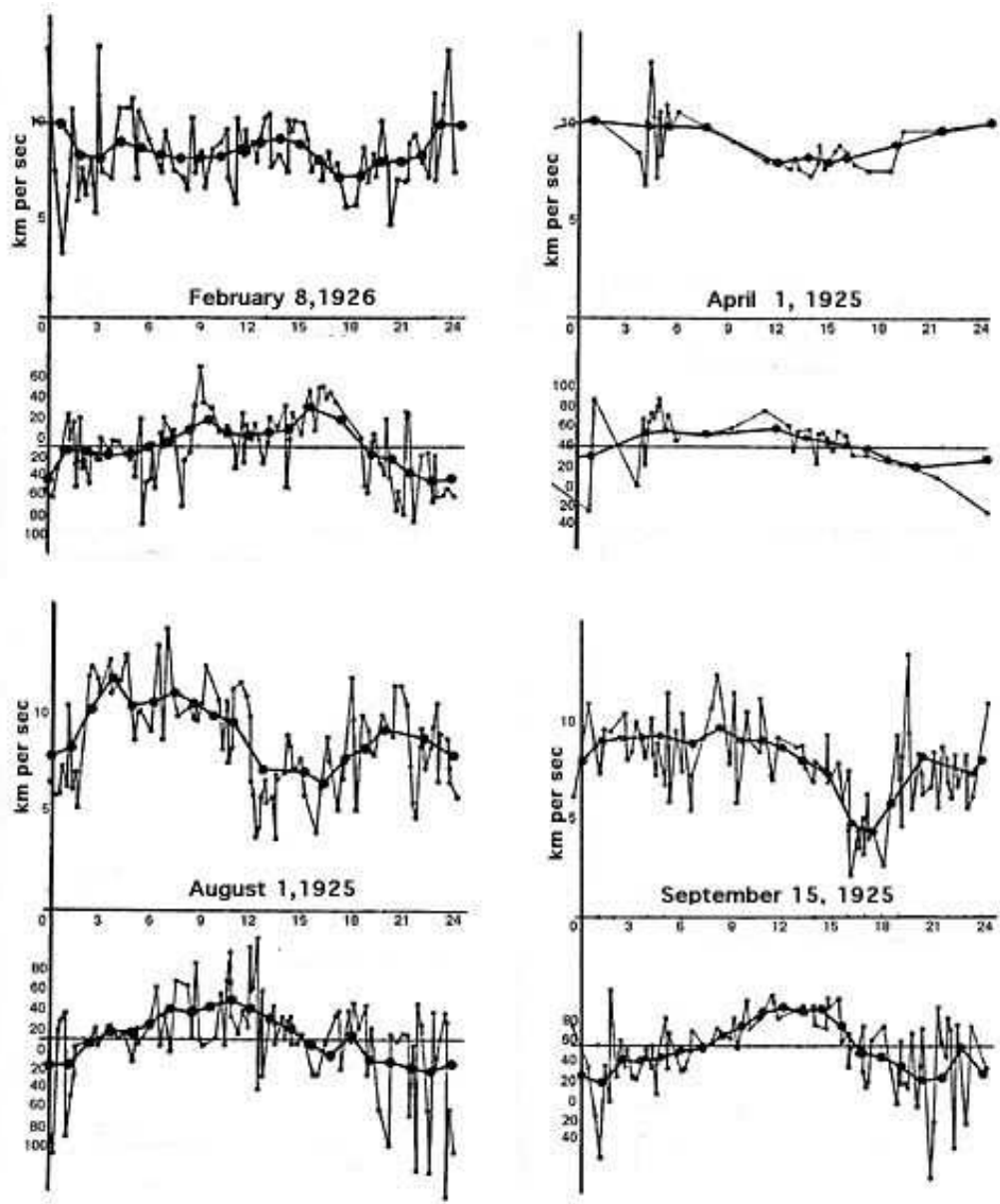}
\caption{\small{ Miller's results from the 1925-1926 observations of absolute motion showing the
 projected `Michelson' speed $v_M$ in km/s  and  azimuth angle $\psi$ in degrees plotted against sidereal time
in hours. The  results are for April, August and September 1925 and February 1926. In most
cases the results arise from observations extending over much of each month, i.e not from a single day 
in each month. Therefore the data points are not strictly in chronological order. The lines joining the
data points are merely to make the data points clearer. The smoother line is a running time
average computed by Miller.  The fluctuations in both $v_M$ and $\psi$ appear to be a combination of
apparatus effects and genuine physical phenomena  caused by turbulence in the gravitational in-flow of
space  towards the Sun.  Each data point arises    from analysis of the average of twenty full rotations of
the interferometer. The speed data for September is re-plotted in Fig.\ref{fig:SeptPlot} showing the corrected
absolute speed.}  
\label{fig:MillerData}}\end{figure}

\begin{figure}
\begin{minipage}[t]{60mm}
\hspace{3mm}\includegraphics[scale=0.9]{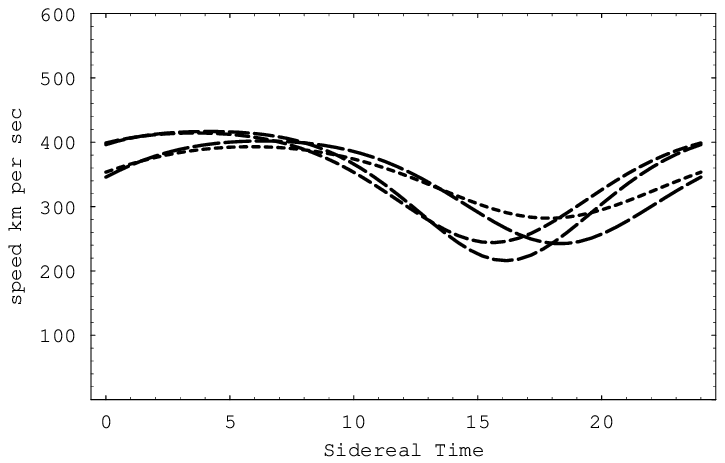}
\makebox[80mm][c]{\small{(a)}}
\end{minipage}
\begin{minipage}[t]{70mm}
\vspace{-42mm}\hspace{10mm}\includegraphics[scale=0.9]{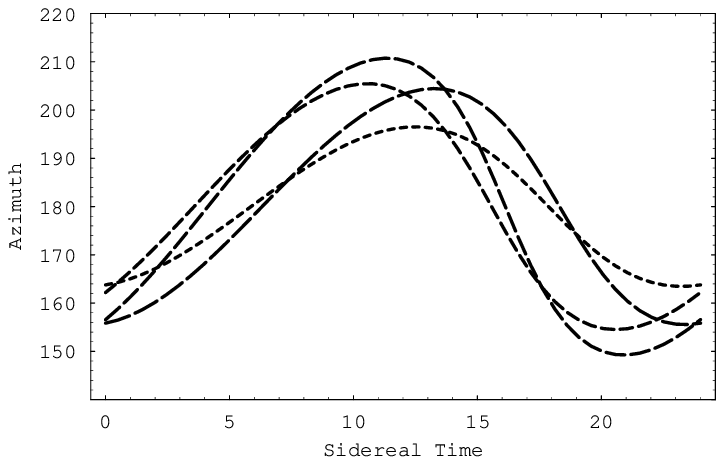}
\makebox[90mm][c]{\small{(b)}}
\end{minipage}
\vspace{0mm}
\caption{\small{ Expected theoretical variation of both (a)
the projected velocity $v_P$,  and (b) the azimuth $\psi$ during one sidereal day in the months of April, August, September and
February, labelled by  increasing dash length. These forms assume a cosmic speed of $417$km/s in the
direction $(\alpha=4^{hr} 54^m, \delta=-70^0 33^\prime)$, and the tangential and in-flows velocities as in
(\ref{eqn:QG6b}). These plots show the characteristics of the signature expected in observations of absolute motion.}  
\label{fig:MSpeedsAzimuths}}\end{figure}

Miller performed this analysis of his data by hand,
and the results for April, August and September 1925 and February 1926 are shown in
Fig.\ref{fig:MillerData}.   The speeds shown are the Michelson speeds $v_M$, and these are easily
corrected for the two neglected effects by dividing these $v_M$ by $k_{air}=\surd{(n^2-1)}=0.0241$, as in
(\ref{eqn:QG6}). Then for example a speed of $v_M=10km/s$ gives $v_P=v_M/k_{air}=415$km/s.  However this correction
procedure was not available to Miller.  He understood that the theory of the Michelson interferometer was
not complete, and so he introduced the phenomenological parameter $k$ in (\ref{eqn:QG1}). We shall denote
his values by  $\overline{k}$. Miller noted, in fact, that
$\overline{k}^2<<1$, as we would now expect.  Miller then proceeded on the assumption  that ${\bf v}$
should have only two components: (i) a cosmic velocity of the Solar system through space, and (ii) the
orbital velocity of the Earth about the Sun. Over a year this vector sum  would result in a changing
${\bf v}$, as was in fact observed, see Fig.\ref{fig:MillerData}.  Further, since the  orbital speed was
known, Miller was able to extract from the data the magnitude and direction of ${\bf v}$ as the orbital
speed offered an absolute scale.  For example the dip in the $v_M$ plots for sidereal times $\tau \approx
16^{hr}$ is a clear indication of the direction of ${\bf v}$, as the dip arises at those sidereal
times when the projection $ v_P$ of  ${\bf v}$ onto the plane of the interferometer is at a minimum. During
a 24hr period the value of $v_P$  varies due to the Earth's rotation.  As well the $v_M$ plots vary
throughout the year because the vectorial sum of the Earth's orbital velocity ${\bf v}_{tangent}$ and the
cosmic velocity  ${\bf v}_{cosmic}$ changes.  There are two effects here as the direction of 
${\bf v}_{tangent}$ is determined by both the yearly progression of the Earth in its orbit about the
Sun, and also because  the plane of the ecliptic is inclined at $23.5^0$ to the celestial plane. 
Figs.\ref{fig:MSpeedsAzimuths}  show the expected theoretical variation of both
$v_P$ and the azimuth $\psi$  during one sidereal day in the months of April, August, September and
February. These plots show the clear  signature of absolute motion effects as seen in the actual
interferometer data of Fig.\ref{fig:MillerData}.

Note that the above corrected Miller projected absolute  speed of
approximately  $v_P=415$km/s  is completely consistent with the corrected   projected
absolute speed of some $330$km/s from the Michelson-Morley experiment, though neither Michelson nor Miller were
able to apply this correction.  The difference in magnitude is completely explained by  Cleveland having a
higher latitude than Mt. Wilson, and also by the only two sidereal times of the Michelson-Morley observations. 
So from his 1925-1926 observations Miller had completely confirmed the true validity of the Michelson-Morley
observations and was able to conclude, contrary to their published conclusions, that the 1887 experiment
had in fact detected absolute motion.  But it was too late. By then the  physicists had incorrectly come to
believe that absolute motion was  inconsistent with various `relativistic effects' that had by then been
observed. This was because the Einstein formalism had been `derived'  from the assumption that
absolute motion was  without meaning and so unobservable in principle. Of course the earlier interpretation  of
relativistic effects by Lorentz had by then lost out to the Einstein interpretation.

\begin{figure}[ht]
\hspace{10mm}\includegraphics[scale=1.7]{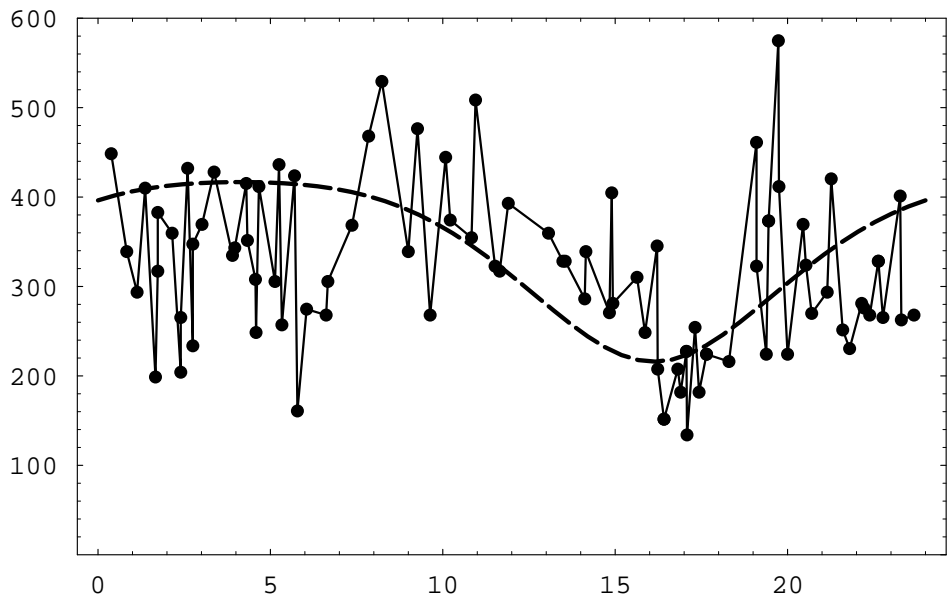}
\caption{\small{  The absolute projected speeds $v_P$,  plotted against sidereal time in hours, 
for September 1925 is re-plotted  using the correction
$v_P=v_M/k_{air}$. The data points were obtained from  Miller's original data which was found in
2002. The dip at $\tau \approx 16^{hr}$(ST) is also seen in the New Bedford data,
Fig.\ref{fig:NewBedford}, and the DeWitte data, Fig.\ref{fig:DeWittetimes}.  So three different 
experiments reveal the same absolute motion signature. The fluctuations in  $v_P$  appear to be
a combination of experimental effects and genuine physical phenomena  caused by turbulence in the
gravitational in-flow of space  towards the Sun.}\label{fig:SeptPlot}}\end{figure}

\subsection{    Gravitational In-flow from the Miller Data\label{subsection:gravitationalinflow}}

As already noted Miller was led to the conclusion that for reasons unknown  the existing theory of
the Michelson interferometer did not reveal true values of
$v_P$, and for this reason  he introduced the parameter $k$,  with $\overline{k}$ indicating his numerical
values. Miller had reasoned  that he could determine both ${\bf v}_{cosmic}$ and $\overline{k}$ by observing the
interferometer determined $v_P$ and $\psi$ over a year because the known orbital velocity  of the Earth about
the Sun would modulate both of these observables, and by a scaling argument he could determine the absolute
velocity of the Solar system.   In this manner he finally determined that $|{\bf v}_{cosmic}|=208$ km/s in the
direction $(\alpha=4^{hr} 54^m, \delta=-70^0 33^\prime)$.  However now that the theory of the Michelson
interferometer has been revealed an anomaly becomes apparent.  Table 3 shows $v=v_M/k_{air}$ 
for each of the four epochs, giving speeds consistent with the revised Michelson-Morley data.  However Table
3 also shows that $\overline{k}$ and the speeds $\overline{v}=v_M/\overline{k}$ determined by the
scaling argument are considerably different.  Here the $v_M$ values arise after taking account of 
the projection effect. That 
$\overline{k}$ is considerably larger than the value of
$k_{air}$ indicates that  another velocity component has been overlooked.   Miller of course only knew of
the tangential  orbital speed of the Earth, whereas the new physics predicts that as-well there is a
quantum-gravity radial in-flow ${\bf v}_{in}$ of the quantum foam. We can re-analyse  Miller's data to
extract a first approximation to the speed of this in-flow component.   Clearly it is
$v_R=\sqrt{v_{in}^2+v^2_{tangent}}$ that sets the scale and not
$v_{tangent}$, and because
$\overline{k}=v_M/v_{tangent}$ and $k_{air}=v_M/v_R$ are the scaling relations, then 
\begin{eqnarray}\label{eqn:QG9}
v_{in}&=&v_{tangent}\sqrt{\displaystyle{ \frac{v_R^2}{v_{tangent}^2}-1 }},  \nonumber \\
      &=&v_{tangent}\sqrt{\displaystyle{ \frac{\overline{k}^2}{k_{air}^2}-1 }}.  
\end{eqnarray} 
\vspace{3mm}

\begin{figure}
\footnotesize{
\hspace{4mm}\begin{tabular}{|l|l|c|l|l|l|l|} 
\hline 
{\bf Epoch} &\mbox{\ \ }$v_M$ & $\overline{k}$ &$ v=v_M/k_{air}$  & $\overline v=v_M/\overline{k}$ 
&$v=\sqrt{3}\overline{v}$&\mbox{\ \  }
$v_{in}$\\
\hline\hline  
 February 8  &9.3 km/s & 0.048 & 385.9 km/s  & 193.8 km/s & 335.7 km/s &  51.7 km/s \\ \hline
 April 1  & 10.1 &0.051 &419.1  & 198.0 &342.9  &56.0 \\ \hline 
 August 1  & 11.2 &0.053 &464.7  & 211.3 &366.0  &58.8 \\ \hline
 September 15  & 9.6 &0.046 &398.3  & 208.7 &361.5  & 48.8\\ \hline
\hline
\end{tabular}}
\vspace{2mm}

\hspace{7mm}  {\small Table 3. The $\overline{k}$ anomaly,  $\overline{k} \gg k_{air}=0.0241$, as  the
gravitational in-flow effect.
  
\hspace{7mm}  Here $v_M$ and $\overline{k}$ come from fitting the interferometer data, while $\,v\,$ and
$\,\overline{v}\,$  are

\hspace{7mm}  computed speeds using the indicated scaling. The average of the in-flow speeds

\hspace{7mm}  is $v_{in}=54\pm5$ km/s, compared to the `Newtonian' in-flow speed of $42$ km/s.

\hspace{7mm} From column 4 we obtain the average $v=417\pm40$km/s. } 
\end{figure}

 Using the  $\overline{k}$  values in Table 3 and the value\footnote{We have not modified this
value to take account of the altitude effect  or temperatures atop  Mt.Wilson. This weather information was not
recorded by Miller. The temperature and pressure effect  is that $n=1.0+0.00029\frac{P}{P_0}\frac{T_0}{T}$, where $T$
is the temperature in $^0$K and
$P$ is the pressure in atmospheres.  $T_0=273$K  and
$P_0=$1atm.} of
$k_{air}$  we obtain the
$v_{in}$ speeds  shown in Table 3, which give an average speed of $54$ km/s, compared to the `Newtonian'
in-flow speed of $42$ km/s.  Note that the in-flow interpretation of the anomaly predicts that
$\overline{k}=(v_R/v_{tangent})\,k_{air}=\sqrt{3}\, k_{air}=0.042$. Of course this simple re-scaling of the
Miller results is not completely valid because (i) the direction of ${\bf v}_R$ is of course different to that
of ${\bf v}_{tangent}$, and also not necessarily orthogonal to ${\bf v}_{tangent}$ because of turbulence, and
(ii) also because of turbulence we would  expect some contribution from the in-flow effect of the Earth itself,
namely that it is not always perpendicular to the Earth's surface, and so would give a contribution to a
horizontally operated  interferometer.

An analysis that properly searches for the in-flow velocity
effect clearly requires a complete re-analysis of the Miller data, and this is now possible and underway at
Flinders University as the original data sheets have been found.  It should be noted that the direction
diametrically opposite 
$(\alpha=4^{hr} 54^m, \delta=-70^0 33^\prime)$, namely $(\alpha=17^{hr}, \delta=+68^\prime)$  was at one
stage considered by Miller as being  possible. This is because the Michelson interferometer, being a
2nd-order device, has a directional ambiguity which can only be resolved by using the diurnal motion of the
Earth. However as Miller did not include the in-flow velocity effect in his analysis it is possible that a
re-analysis might give this northerly direction as the direction of absolute motion of the Solar system.

Hence not only did Miller observe absolute motion, as he claimed,  but the quality and quantity of his data
has also enabled the confirmation of  the existence of the gravitational in-flow effect \cite{RC03, RC05}.  This
is a manifestation of a new theory of gravity and one which relates to quantum gravitational effects via the
unification of matter and space developed in previous sections. As well the persistent evidence that this
in-flow is turbulent indicates that this theory of gravity involves self-interaction of space itself.

\subsection{  The Illingworth Experiment: 1927\label{subsection:theillingworth}}

\begin{figure}[ht]
\hspace{15mm}\includegraphics[scale=1.6]{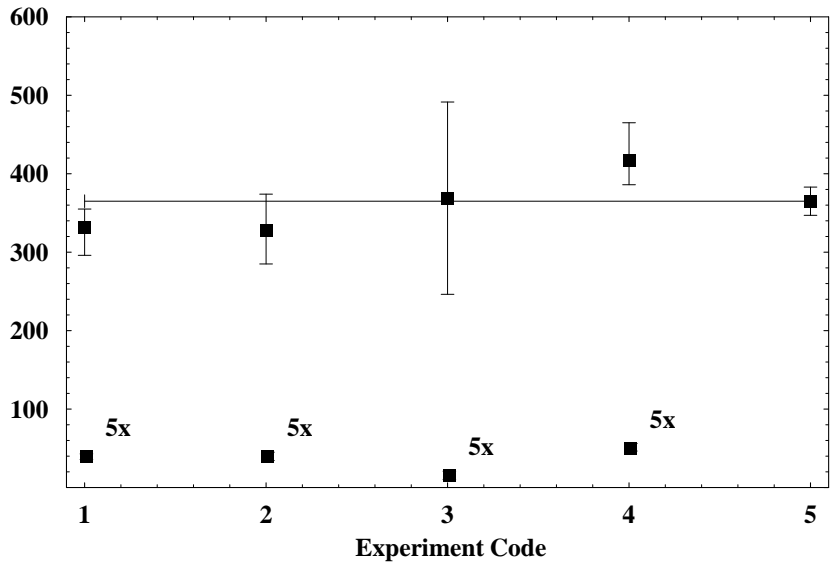}
\caption{\small 
Speeds $v$ in km/s determined from various Michelson interferometer experiments ({\bf 1})-({\bf 4}) and CMB ({\bf
5}): ({\bf 1}) Michelson-Morley (noon observations) and  ({\bf 2})
 ($18^h$ observations) see Sect.\ref{subsection:themichelsonmorley}, ({\bf 3}) Illingworth  
\cite{Illingworth}, ({\bf 4}) Miller, Mt.Wilson
\cite{Miller2}, and finally in ({\bf 5}) the speed from  observations of the CMB
 spectrum dipole term
\cite{CMB}.  The results ({\bf 1})-({\bf 3}) are not
corrected for the
$\pm30$km/s of the orbital motion of the Earth about the Sun or for the gravitational in-flow speed, though
these correction were made for ({\bf 4}) with the speeds from Table 3.  The horizontal line at
$v=369$km/s is to aid comparisons with the CMB frame speed data.  The Miller direction is different to the CMB
direction.  Due to the angle between the velocity vector and the plane of interferometer the results ({\bf
1})-({\bf 3})  are less than or equal to the true speed, while the result for ({\bf 4}) is the true speed as
this projection effect was included in the analysis.  These results demonstrate the remarkable consistency
between the three  interferometer experiments. The Miller speed  agrees with the speed from the DeWitte
non-interferometer experiment, in Sect.\ref{subsection:dewitte}. The lower data, magnified by a factor of
5, are the original speeds $v_{M}$  determined from fringe shifts using (\ref{eqn:QG0}) with
$k=1$. This figure updates the corresponding figure in Ref.\cite{CK}.  } 
\label{fig:AllSpeeds}\end{figure}

In 1927 Illingworth \cite{Illingworth} performed a Michelson interferometer experiment in which the light
beams passed through the gas  Helium,  
\begin{quote}{\it ...as it has such a low index of refraction that variations due to temperature changes are
reduced to a negligible quantity.}
\end{quote}
For Helium at STP $n=1.000036$ and so $k^2_{He}=0.00007$, which results in an enormous reduction in sensitivity
of the interferometer.  Nevertheless this experiment gives an excellent opportunity to check the
$n$ dependence in (\ref{eqn:QG6}).   Illingworth, not surprisingly, reported no ``ether drift to an accuracy
of  about one kilometer per second''.  M\'{u}nera \cite{Munera} re-analysed the Illingworth data to obtain a
speed  $v_M=3.13 \pm 1.04$km/s.  The correction  factor in (\ref{eqn:QG6}),
$1/\sqrt{n_{He}^2-1}=118$, is large for Helium and gives $v=368\pm 123$km/s. As shown in 
Fig.\ref{fig:AllSpeeds} the Illingworth observations now agree  with those of Michelson-Morley  and  Miller,
though they would certainly be inconsistent without the $n-$dependent correction, as shown in the lower data
points (shown at $5\times$ scale). 

So the use by Illingworth of Helium gas has turned out have offered a fortuitous opportunity to confirm the
validity of the refractive index effect, though because of the insensitivity of this experiment the resulting
error range is significantly larger than those of the other interferometer  observations.  So finally it is
seen that the Illingworth experiment detected absolute motion with a speed consistent with all other
observations.   

\subsection{  The New Bedford Experiment: 1963\label{subsection:thenewbedford}}

\begin{figure}
\hspace{5mm}\includegraphics[scale=1.0]{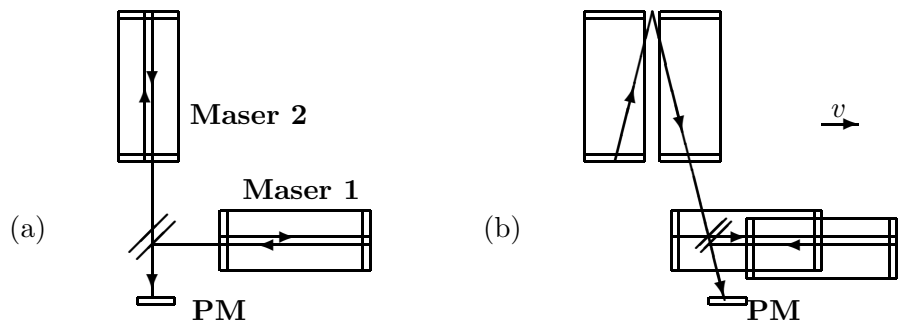}
\vspace{29mm}
\caption{\small{ Schematic diagram for recording the variations in beat frequency between two optical masers:
 (a) when at absolute rest, (b) when in absolute motion at velocity ${\bf v}$. PM is the photomultiplier
detector. The apparatus was rotated back and forth through $90^0$.}  
\label{fig:Masers}}\end{figure}

\begin{figure}[h]
\hspace{10mm}\includegraphics[scale=1.5]{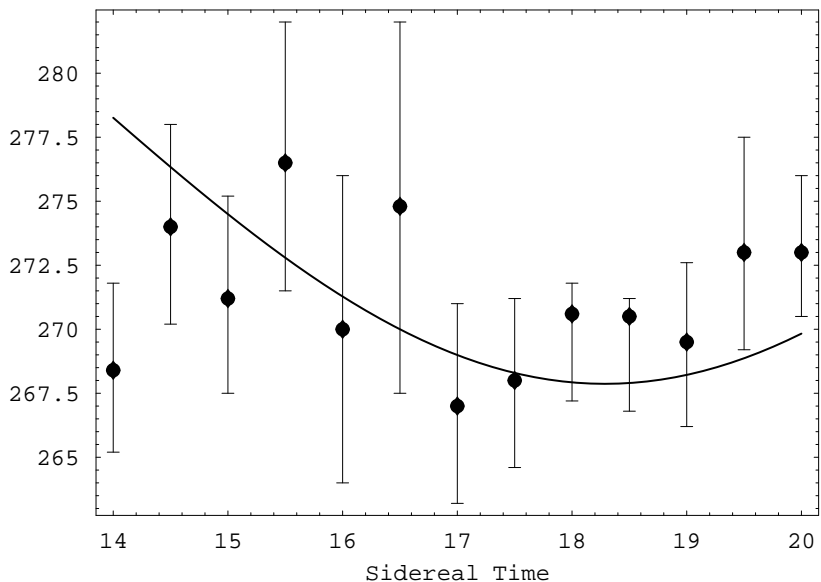}
\caption{\small{  Frequency difference in kHz between the two masers in the 1963 New Bedford experiment after
a $90^0$ rotation. The $275$kHz difference is a systematic repeatable apparatus effect, whereas the superimposed
`dip' at
$17-18\!\!:\!\!00^{hr}$ sidereal time of approximately $3$kHz is a real  time dependent frequency difference. 
The full curve shows the theoretical prediction for the time of the `dip' for this experiment using the Miller
direction for  ${\bf \hat{v}}$ $(\alpha=4^{hr} 54^m, \delta=-70^0 33^\prime)$ with $|{\bf v}|=417$km/s and
including the Earth's orbital velocity  and Sun gravitational in-flow velocity effects for January 20, 1963. The
absolute scale of this theoretical prediction was not possible to compute as the refractive index of the He-Ne
gas mixture was unknown. }  
\label{fig:NewBedford}}\end{figure}

In 1964 from an absolute motion detector  experiment at New Bedford, latitude $42^0$N, Jaseja
{\it et al} \cite{Jaseja} reported yet  another `null result'. In this experiment  two He-Ne masers were
mounted with axes perpendicular on a rotating table, see  Fig.\ref{fig:Masers}. Rotation of the table through
$90^0$ produced repeatable variations in the frequency difference of about $275$kHz, an effect  attributed
to  magnetorestriction in the Invar spacers due to the Earth's magnetic field.  Observations over some
six consecutive hours on January 20, 1963 from $6\!\!:\!\!00$ am to  $12\!\!:\!\!00$ noon local time  did
produce  a `dip' in the frequency difference of some $3$kHz superimposed on the $275$kHz effect, as shown in
Fig.\ref{fig:NewBedford} in which the local times have been converted to sidereal times.  The most noticeable
feature is that the dip occurs at approximately $17-18\!\!:\!\!00^{hr}$ sidereal time (or $9-10\!\!:\!\!00$ hrs
local time), which agrees with the direction of absolute motion observed by Miller and also by DeWitte (see
Sect.\ref{subsection:dewitte}). It was most fortunate that this particular time period was chosen as at other
times the effect is much smaller, as shown for example in Fig.\ref{fig:SeptPlot} for August or more
appropriately for the February data in Fig.\ref{fig:MillerData} which shows the minimum at $18\!\!:\!\!00^{hr}$
sidereal time. The local times were chosen by Jaseja {\it et al} such that if the only motion was due to the
Earth's orbital speed the maximum frequency difference, on rotation,    should have occurred at $12\!\!:\!\!00$hr
local time, and the minimum frequency difference at $6\!\!:\!\!00$ hr local time, whereas in fact the minimum
frequency difference occurred at $9\!\!:\!\!00$ hr local time.

As for the Michelson-Morley experiment the analysis of the New Bedford experiment was also bungled. Again this
apparatus can only detect the  effects of absolute motion if the   cancellation between  the
geometrical effects and Fitzgerald-Lorentz length contraction effects is incomplete as  occurs only when the
radiation travels in a gas, here the He-Ne gas present in the maser.  

This double maser apparatus is essentially
equivalent to a Michelson interferometer.   Then the resonant frequency $\nu$ of each maser is
proportional to the reciprocal of the out-and-back travel time. For  maser 1
\begin{equation}\label{eqn:Maser1}
\nu_1=m\frac{V^2-v^2}{2LV\sqrt{1-\displaystyle\frac{v^2}{c^2}}},
\end{equation}
for which a Fitzgerald-Lorentz contraction occurs, while for maser 2
\begin{equation}\label{eqn:Maser2}
\nu_2=m\frac{\sqrt{V^2-v^2}}{2L}.
\end{equation}
Here $m$ refers to the mode number of the masers. When the apparatus is rotated the net observed 
frequency difference is
$\delta
\nu =2(\nu_2-\nu_1)$, where the factor of `2' arises  as the roles of the two masers are reversed after a
$90^0$ rotation. Putting $V=c/n$  we find for $v << V$  and with $\nu_0$  the at-rest resonant frequency,  that
\begin{equation}\label{eqn:Maser3}
 \delta \nu=(n^2-1)\nu_0\frac{v^2}{c^2}+O(\frac{v^4}{c^4}).
\end{equation}
 If we use the Newtonian physics analysis, as in Jaseja {\it et al} \cite{Jaseja}, which neglects both the
Fitzgerald-Lorentz contraction and the refractive index effect, then we obtain  $\delta \nu=\nu_0 v^2/c^2$,
that is without the $n^2-1$ term, just as for the Newtonian analysis of the Michelson interferometer itself.  Of
course the very small magnitude of the absolute motion effect, which was approximately 1/1000 that expected
assuming only  an orbital speed of $v=30$ km/s in the Newtonian analysis,    occurs simply because the
refractive index of the He-Ne gas is very close to one\footnote{It is possible to compare the 
refractive index of  the He-Ne gas mixture in the maser  with the value extractable from this data:
$n^2=1+30^2/(1000\times 400^2)$, or $n=1.0000028$.}. Nevertheless given that it is small  the sidereal time of
the  obvious 'dip' coincides almost exactly with that of the other observations of absolute motion. 
  
 The New Bedford experiment was yet another missed opportunity to have revealed the existence of absolute
motion. Again the spurious argument was that because the  Newtonian physics analysis gave the wrong prediction
then Einstein relativity must be correct.  But the analysis simply failed to take account of  the
Fitzgerald-Lorentz contraction, which had been known since the end of the 19$^{th}$ century,  and  the
refractive index effect which had an even longer history.  As well the authors failed to convert their
local times to sidereal times and compare the time for the `dip' with Miller's time\footnote{There is no
reference to Miller's 1933 paper in  Ref.\cite{Jaseja}. }.

\subsection{  The DeWitte Experiment: 1991\label{subsection:dewitte}}

The Michelson-Morley, Illingworth, Miller and New Bedford experiments  all used Michelson
interferometers or its equivalent in gas mode, and all revealed  absolute motion.  The Michelson
interferometer is a 2nd-order device meaning that the time difference between the `arms' is proportional to
$(v/c)^2$.  There is also a factor of  $n^2-1$ and for gases like air and particularly
Helium  or Helium-Neon mixes this results in very small time differences and so these experiments
were always very difficult.  Of course without the gas the Michelson interferometer is
incapable of detecting absolute motion\footnote{So why not use a transparent solid in
place of the gas? See Sect.\ref{subsection:solidstate} for the discussion.}, and so there
are fundamental limitations to the use of this interferometer in the study of absolute
motion and related effects. 
 
\begin{figure}[t]
\hspace{15mm}\includegraphics[scale=1.5]{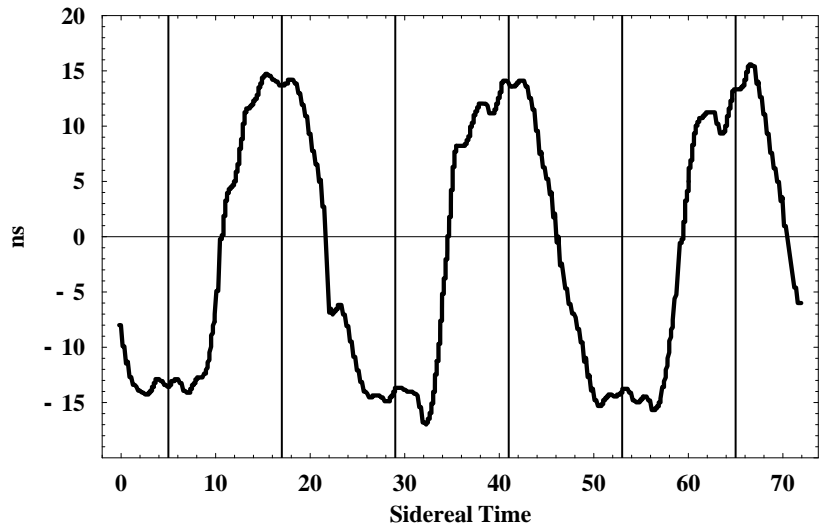}
\caption{\small{ Variations in twice the one-way travel time, in ns, for an RF signal to travel 1.5
km through a coaxial cable between  Rue du Marais and Rue de la Paille, Brussels.  An offset  has been used 
such that the average is zero.  The definition of the sign  convention for $\Delta t$ used by DeWitte
is unclear.  The cable has a
North-South  orientation, and the data is $\pm$ difference of the travel times  for NS and SN
propagation.  The sidereal time for maximum  effect of $\sim\!\!17$hr (or   $\sim\!\!5$hr) (indicated
by vertical lines) agrees with the direction found by Miller and also by  Jaseja {\it et al}, but because of the
ambiguity in the definition of $\Delta t$ the opposite direction would also be consistent  with this data. Plot shows
data over 3 sidereal days  and is plotted against sidereal time. See Fig.\ref{fig:DeWitteTheory}b for
theoretical predictions for one sidereal day. The time of  the year of the data is not identified.
 The fluctuations are evidence of turbulence associated
with  the gravitational in-flow towards the Sun. Adapted from DeWitte \cite{DeWitte}.}  
\label{fig:DeWittetimes}}\end{figure}

In a remarkable development  in 1991 a research project  within Belgacom,
the Belgium telecommunications company, stumbled across yet another detection of absolute motion, and one
which turned out to be 1st-order in $v/c$.  The study was undertaken by  Roland DeWitte
\cite{DeWitte}.   This organisation had two sets of atomic clocks in two buildings in Brussels separated by
1.5 km and the research project  was an investigation of  the task of synchronising these two clusters of
atomic clocks. To that end  5MHz radiofrequency signals were sent  in both directions   through two  buried 
coaxial cables linking the two clusters.   The atomic clocks were caesium beam
atomic clocks, and there were three in each cluster. In that way the stability of the clocks could
be established and monitored. One cluster was in a building on Rue du Marais and the second cluster
was due south in a building on Rue de la Paille.  Digital phase comparators were used to measure
changes in times between clocks within the same cluster and also in the propagation times of the RF
signals. Time differences between clocks within the same cluster showed  a linear phase drift caused
by the clocks not having exactly the same frequency together with short term and long term noise.
However the long term drift was very linear and reproducible, and that drift could be allowed for
in analysing time differences in the propagation times between the clusters.

Changes in propagation times  were observed and eventually observations over  178 days were recorded. A sample
of the  data, plotted against sidereal time for just  three days, is shown in Fig.\ref{fig:DeWittetimes}. 
DeWitte recognised that the data was evidence of absolute motion but he was unaware of the Miller experiment 
and did not realise that the Right Ascension for maximum/minimum  propagation time agreed almost
exactly with Miller's direction $(\alpha, \delta)=(17.5^h, 65^0)$. In fact DeWitte expected that the direction
of absolute motion should have been in the CMB direction, but that would have given the data a totally
different sidereal time signature, namely the times for maximum/minimum would have been shifted by 6 hrs. 
The declination of the velocity observed in this DeWitte experiment cannot be determined from the data as
only three days of data are available.   However assuming exactly the same declination as Miller  the speed
observed by DeWitte appears to be also  in excellent agreement with the Miller speed, which in turn is in
agreement with that from the Michelson-Morley and Illingworth experiments, as shown in
Fig.\ref{fig:AllSpeeds}.  

\begin{figure}[ht]
\begin{minipage}[t]{60mm}
\hspace{2mm}\includegraphics[scale=0.9]{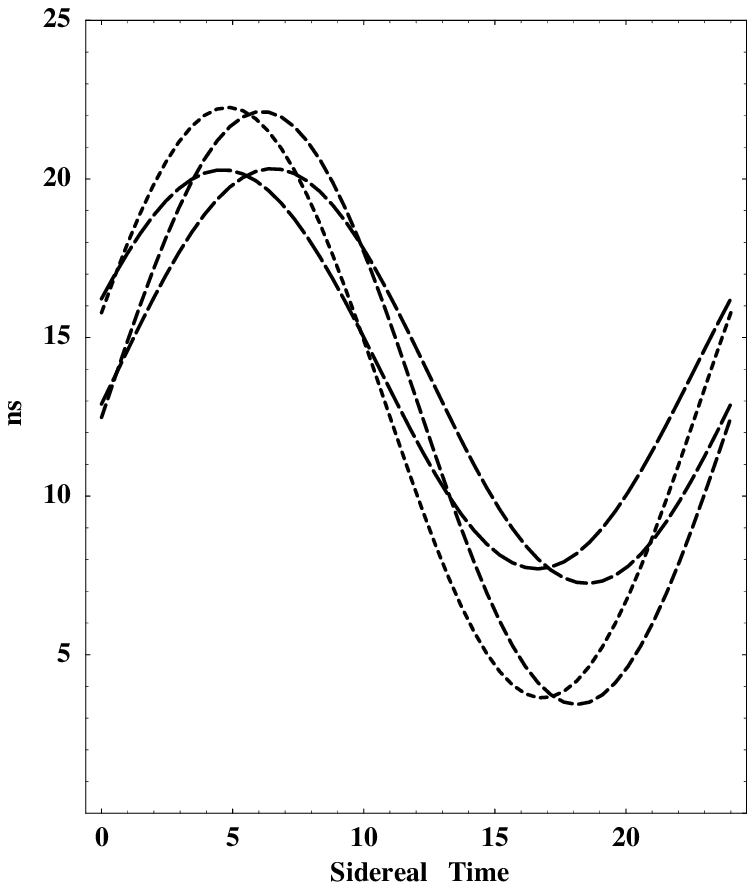}
\makebox[80mm][c]{\small{(a)}}
\end{minipage}
\begin{minipage}[t]{70mm}
\vspace{-79mm}\hspace{15mm}\includegraphics[scale=0.9]{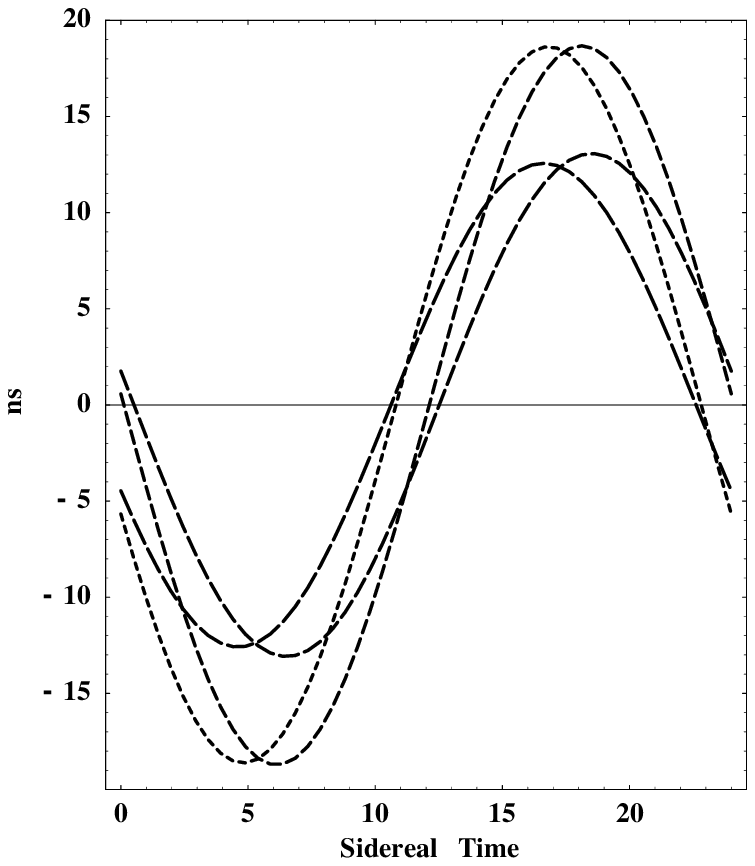}
\makebox[110mm][c]{\small{(b)}}
\end{minipage}
\vspace{0mm}
\caption{\small{ Theoretical predictions for the variations in travel time,  in ns, for
one sidereal day, in the DeWitte Brussels coaxial cable experiment  for ${\bf v}_{cosmic}$ in  the direction
$(\alpha, \delta)=(17.5^h, 65^0)$ and with the Miller magnitude of 417 km/s, and including  orbital and
in-flow effects (but without turbulence). Shown are the results for four days: for the Vernal Equinox,  March
21 (shortest dashes), and for 90, 180 and 270 days later (shown with increasing dash length). 
Figure (a) Shows change in one-way travel time $t_0nv_P/c$ for signal travelling from N to S.  Figure (b)
shows  $\Delta t$, as defined in (\ref{eqn:DW1}), with an offset such that  the average is zero so as to
enable comparison with the  data in Fig.\ref{fig:DeWittetimes}. $\Delta t$ is twice the one-way travel
time. For the direction opposite to
$(\alpha, \delta)=(17.5^h, 65^0)$ the same curves arise except that the identification of the months is
different and  the sign of $\Delta t$ also changes. The sign of $\Delta t$  determines which of the two
directions is the actual direction of absolute motion. However the definition of the sign  convention for
$\Delta t$ used by DeWitte is unclear.}
\label{fig:DeWitteTheory}}
\end{figure} 

Being  1st-order in $v/c$  the Belgacom experiment is easily analysed to sufficient accuracy by
ignoring relativistic effects, which are 2nd-order in $v/c$.   Let the projection of the absolute
velocity vector ${\bf v}$ onto the direction of the coaxial cable be $v_P$ as before.  Then the
phase comparators reveal the difference  between the propagation
times in NS and SN directions. First consider the analysis with no  Fresnel drag effect,
\begin{eqnarray}
\Delta t &=& \frac{L}{\displaystyle{\frac{c}{n}}-v_P}-
\frac{L}{\displaystyle{\frac{c}{n}}+v_P},\nonumber\\
&=& 2\frac{L}{c/n}n\frac{v_P}{c}+O(\frac{v_P^2}{c^2}) \approx 2t_0n\frac{v_P}{c}.
\label{eqn:DW1}\end{eqnarray}

Here $L=1.5$ km is the length of the coaxial cable, $n=1.5$ is the refractive index of the insulator within
the coaxial cable, so that the speed of the RF signals is approximately $c/n=200,000$km/s, and so
$t_0=nL/c=7.5\times 10^{-6}$ sec is the one-way RF travel time when
$v_P=0$.  Then, for example, a  value of  $v_P=400$km/s would give $\Delta t = 30$ns.  Because Brussels has a
latitude of $51^0$ N then for the Miller direction the projection effect is such that $v_P$ almost varies
from zero to a maximum value of $|{\bf v}|$.  The DeWitte  data in  Fig.\ref{fig:DeWittetimes}
shows $\Delta t$ plotted with a false zero, but  shows a variation of some 28 ns.  So the DeWitte
data is in excellent  agreement with the Miller's data\footnote{There is ambiguity in
Ref.\cite{DeWitte} as to whether the   time variations  in  Fig.\ref{fig:DeWittetimes}  include the
factor of 2 or not, as defined in (\ref{eqn:DW1}). It is assumed here that a factor of 2 is included. }.
The Miller experiment has thus been confirmed by a non-interferometer experiment if we ignore a Fresnel
drag.

 But if we include a Fresnel drag effect then the change in travel time $\Delta t_F$ 
becomes
\begin{eqnarray}
\Delta t_F &=& \frac{L}{\displaystyle{\frac{c}{n}}+bv_P-v_P}-
\frac{L}{\displaystyle{\frac{c}{n}}-bv_P+v_P},\nonumber\\
&=& 2\frac{L}{c}\frac{v_P}{c}+O(\frac{v_P^2}{c^2}) ,\nonumber \\
&=&  \frac{1}{n^2}\Delta t,
\label{eqn:DWFD}\end{eqnarray}
where $b=1-1/n^2$ is the Fresnel drag coefficient. Then $\Delta t_F$
is smaller than
$\Delta t$ by a factor of $n^2=1.5^2=2.25$, and so a speed of $v_P=2.25\times 400=900$ km/s would be
required to produce a $\Delta t_F= 30 $ ns. This speed is inconsistent with the results from gas-mode
interferometer experiments, and also inconsistent with the data from the Torr-Kolen gas-mode coaxial cable
experiment, Sect.\ref{subsection:Thetorr}.  This raises the question as to whether the Fresnel effect is
present in
 transparent solids, and indeed whether it has ever been studied? As well we are assuming  the
conventional eletromagnetic theory for the RF fields in the coaxial cable.  An experiment to investigate this is
underway at Flinders university \cite{RCPP2003}.

The actual days of the data in 
Fig.\ref{fig:DeWittetimes} are not revealed in Ref.\cite{DeWitte} so a detailed analysis of the DeWitte data
is not possible. Nevertheless theoretical predictions for various days in a year are shown in
Fig.\ref{fig:DeWitteTheory} using the Miller speed of $v_{cosmic}=417$ km/s (from Table 3) and  where the
diurnal effects of the Earth's orbital velocity and the gravitational in-flow cause the range of variation of
$\Delta t$ and sidereal  time of maximum effect to vary throughout the year. The predictions give $\Delta t
= 30\pm 4$ ns over a year compared to the  DeWitte value of 28 ns in Fig.\ref{fig:DeWittetimes}.  If
all of DeWitte's 178 days of data were available then a detailed analysis would be possible. 

\begin{figure}[t]
\hspace{15mm}\includegraphics[scale=1.5]{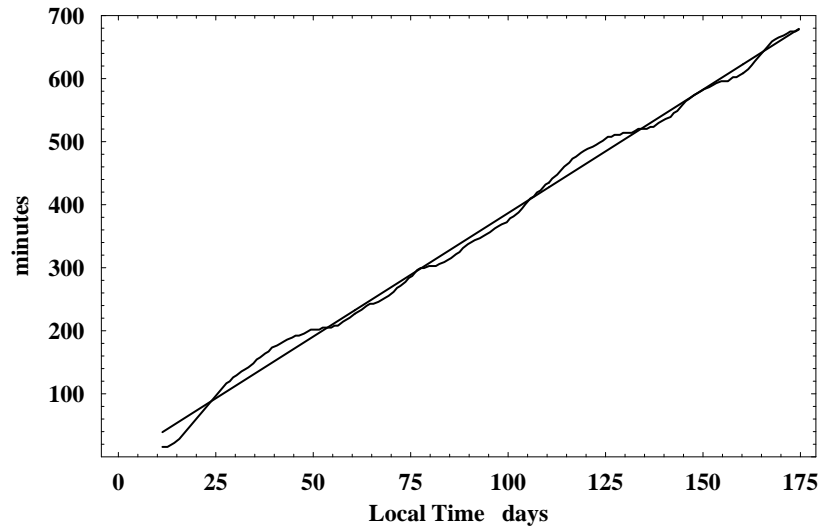}
\caption{\small{ Plot of the negative of the drift of the cross-over time between minimum and
maximum travel-time variation each day (at $\sim10^h\pm1^h$ ST) versus local solar time for some
180 days. The straight line plot is the least squares fit to the experimental data, 
 giving an average slope of 3.92 minutes/day. The time difference between a sidereal day and a solar
day is 3.93 minutes/day.    This demonstrates that the effect is related to sidereal time and not local
solar time.   The actual days of the year are not identified in Ref.\cite{DeWitte}.
 Adapted from DeWitte \cite{DeWitte}.}  
\label{fig:DeWitteST}}\end{figure}

Ref.\cite{DeWitte} does however reveal the sidereal time of the cross-over time, that is a `zero' time
in Fig.\ref{fig:DeWittetimes}, for all 178 days of data.  This is plotted in Fig.\ref{fig:DeWitteST} and
demonstrates that the time variations are correlated with sidereal time and not local solar time.  A
least squares best fit of a linear relation to that data gives that the cross-over time is retarded, on
average, by 3.92 minutes per solar day. This is to be compared with the fact that a sidereal day is 3.93 minutes
shorter than a solar day. So the effect is certainly  cosmological and not associated with any daily thermal
effects, which in any case would be very small as the cable is buried.  Miller had also compared his
data against sidereal time and established the same property, namely that up to  small diurnal effects 
identifiable with the  Earth's orbital  motion,  features in the data tracked sidereal time and not
solar time; see Ref.\cite{Miller2} for a detailed analysis.

The DeWitte data is also capable of resolving the question of the absolute direction of motion found by
Miller. Is the direction  $(\alpha, \delta)=(17.5^h, 65^0)$ or the opposite direction? Being a 2nd-order
Michelson interferometer experiment Miller had to rely on the Earth's diurnal effects in order to  resolve
this ambiguity, but his analysis of course did not take account of the gravitational in-flow effect, and so
until a re-analysis of his data his preferred choice of direction must remain to be confirmed.  The DeWitte
experiment could easily resolve this ambiguity by simply noting the sign of $\Delta t$.  Unfortunately it is
unclear  in Ref.\cite{DeWitte}  as to how the sign in Fig.\ref{fig:DeWittetimes} is actually defined, and
DeWitte does not report a direction expecting, as he did, that the direction should have been the same as the
CMB direction.

The DeWitte observations were truly remarkable considering that initially they were serendipitous.  They
demonstrated yet again that the Einstein postulates were in contradiction with experiment.  To my knowledge
no physics journal has published a report of
the  DeWitte experiment. 

That the DeWitte experiment is not a gas-mode Michelson interferometer experiment is very significant.  The
 value of the speed of absolute motion revealed by the  DeWitte experiment of some 400 km/s is in agreement
with the speeds revealed by the new analysis of various  Michelson interferometer data  which uses the
recently discovered refractive index effect, see Fig.\ref{fig:AllSpeeds}. Not only was this effect confirmed
by comparing results for different gases, but the re-scaling of the older $v_M$ speeds to $v=v_M/\sqrt{n^2-1}$
speeds resulting from this effect are now confirmed.  A new and much simpler 1st-order experiment  is
discussed in \cite{RCPP2003} which avoids the use of atomic clocks.

\subsection{  The Torr-Kolen Experiment: 1981\label{subsection:Thetorr}}

\begin{figure}
\vspace{-20mm}
\hspace{30mm}\includegraphics[scale=0.729]{TorrData.epsf}

\vspace{0mm}\hspace{28mm}\includegraphics[scale=1.086]{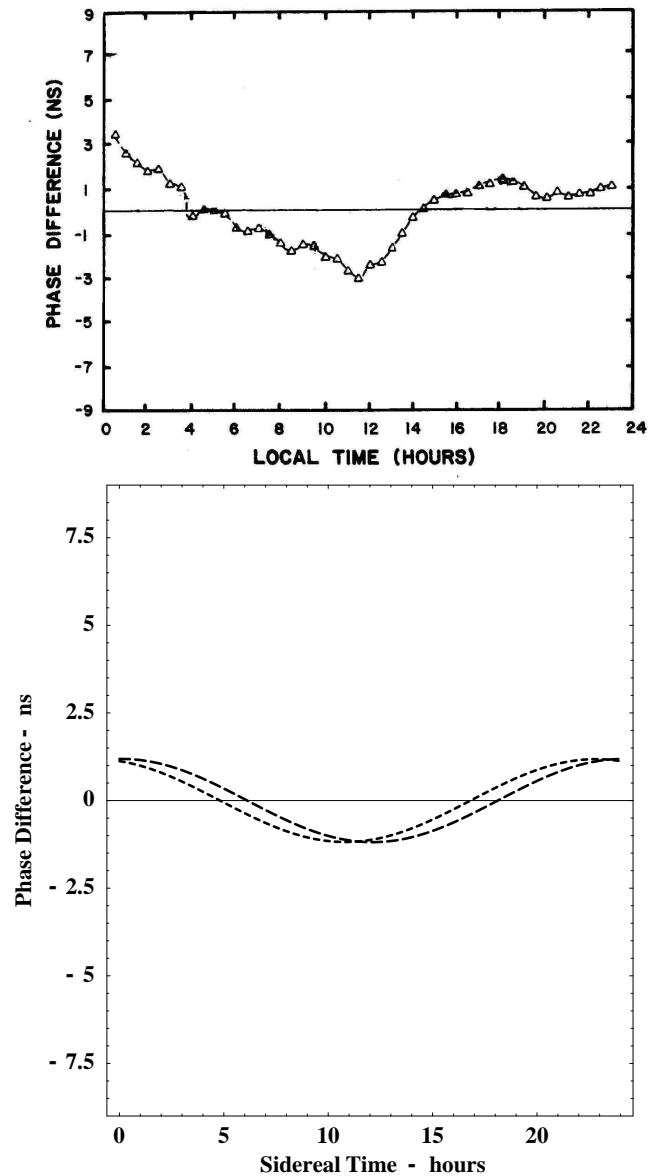}
\caption{\small{ Upper figure is data from the 1981 Torr-Kolen experiment at Logan, Utah \cite{Torr}. 
The data shows variations in travel times
(ns),  for local times,  of an RF signal travelling through 500m of coaxial 
cable  orientated in an E-W direction. Actual days are not indicated but the experiment 
was done during February-June 1981.   Results are for  a typical day.   For the 1st of
February the  local time of $12\!\!:\!\!00$  corresponds to $13\!\!:\!\!00$   sidereal
 time.  The predictions are for   March (shortest dashes) and  June, 
for a cosmic speed of $417$ km/s in the direction  $(\alpha, \delta)=(17.5^h, 65^0)$, and including orbital and in-flow
velocities but without theoretical turbulence.} 
\label{fig:TorrKolen}}
\end{figure} 

A coaxial cable experiment similar to but before the DeWitte experiment was performed at the Utah University in 1981 by Torr and 
Kolen \cite{Torr}. This involved two rubidium vapor clocks placed approximately 500m apart with a 5 MHz sinewave
RF signal propagating between the clocks via a nitrogen filled coaxial cable maintained at a constant pressure
of $\sim$2 psi. This means that the Fresnel drag effect is not important in this experiment. Unfortunately the
cable was orientated in  an East-West direction
which is  not a favourable  orientation for observing  absolute motion in the Miller direction, unlike the 
Brussels North-South cable
orientation.  There is no reference to Miller's result in the Torr and  Kolen paper, otherwise they would presumably not have used
this orientation.  Nevertheless  there is a  projection of the absolute motion velocity onto the
East-West cable and Torr and Kolen did observe an effect in that, while the round speed time remained
constant within 0.0001\%c, typical variations in the one-way travel time  were observed,
as shown in Fig.\ref{fig:TorrKolen} by the triangle data points.  The theoretical  predictions for the 
Torr-Kolen experiment for a cosmic speed of $417$ km/s in the direction  $(\alpha, \delta)=(17.5^h,
65^0)$, and including orbital and in-flow velocities, are  shown in Fig.\ref{fig:TorrKolen}.  As well the maximum effect
occurred, typically,  at the predicted times.  So the results of this experiment are also in remarkable
agreement with the Miller direction, and the speed of 417 km/s which of course only arises after
re-scaling the Miller speeds for the effects of the gravitational in-flow. As well Torr and Kolen
reported fluctuations in both the magnitude  and time of the maximum variations in travel time just as
DeWitte observed some 10 years later.  Again we argue that these fluctuations are evidence of genuine
turbulence in the in-flow as discussed in Sect.\ref{subsection:inflowturbulence}.  So the Torr-Kolen
experiment again shows strong evidence for the new theory of gravity, and which is over and above its
confirmation of the various observations of absolute motion.

\subsection{    Galactic In-flow  and the CMB Frame\label{subsection:galacticinflow}}  
Absolute motion (AM) of the Solar system has been observed in the direction
$(\alpha=17.5^h,\delta=65^0)$, up to an overall sign to be sorted out,  with a speed of
$417 \pm 40$ km/s. This is the velocity after  removing the contribution of the Earth's
orbital speed and the Sun in-flow effect. It is significant that this velocity is different
to that associated with the Cosmic Microwave Background \footnote{The understanding of the galactic 
in-flow effect was not immediate: In
\cite{CK} the direction was not determined, though the speed was found to be comparable to 
the CMB determined speed.  In \cite{RC03} that the directions were very different was 
noted but not appreciated, and in fact thought to be due to experimental error. In
\cite{RC05} an analysis of some of the `smoother' Michelson-Morley data resulted in an 
incorrect direction. At that stage it was not understood that the data showed large 
fluctuations in the azimuth apparently caused by the turbulence. Here the issue is hopefully 
finally resolved.} 
(CMB) relative to which the Solar system
has a speed of $369$ km/s in the direction
 $(\alpha=11.20^h,\delta=-7.22^0)$, see \cite{CMB}. 
This CMB velocity is obtained by finding the preferred frame in which this thermalised
$3^0$K radiation is isotropic, that is by removing the dipole component.  
The CMB velocity is a measure of the motion
of the Solar system relative to the universe as a whole, or aleast a shell of the universe
some 15Gyrs away, and indeed the near uniformity of that radiation in all directions
demonstrates that we may  meaningfully refer to the spatial structure of the
universe.  The concept here is that at the time of decoupling of this radiation from
matter that matter was on the whole, apart from small observable fluctuations, at
rest with respect to the quantum-foam system that is space. So the CMB velocity is the
motion of the Solar system  with respect to space {\it universally},  but not
necessarily with respect to the   {\it local} space.  Contributions to this  velocity
would arise from the orbital motion of the Solar system within the Milky Way galaxy,
which has  a speed of some 250 km/s, and contributions from the motion of the Milky
Way within the local cluster, and so on to perhaps larger clusters.

On the other hand the AM velocity is a vector sum of this {\it universal} CMB
velocity and the net velocity associated with the {\it local} gravitational in-flows into
the Milky Way and the local cluster.  If the CMB velocity had been identical to the AM
velocity then the in-flow  interpretation of gravity would have been proven wrong. We
therefore have three pieces of experimental evidence for this interpretation (i) the
refractive index anomaly discussed previously in connection with the Miller data, (ii) the
turbulence seen in all detections of absolute motion, and now (iii) that the AM velocity is
different in both magnitude and direction from that of the  CMB  velocity, and that this
velocity does not display the turbulence seen in the AM velocity. 

That the AM and CMB velocities are different amounts to the discovery of the resolution 
to the `dark matter' conjecture. Rather than the galactic velocity anomalies being caused by
such undiscovered `dark matter' we see that the in-flow into non spherical galaxies, such as
the spiral Milky Way, will be non Newtonian \cite{RCPP2003}.   As well it will be interesting to determine,
at least theoretically, the scale of turbulence expected in galactic systems,
particularly as the magnitude of the turbulence seen in the AM velocity is somewhat
larger than might be expected from the Sun in-flow alone. Any theory for the turbulence
effect will certainly be checkable within the Solar system as the time scale of this
is suitable for detailed observation.

It is also clear that the time of obervers  at rest with respect to the CMB frame is 
absolute or  universal time.  This interpretation of 
the CMB frame has of course always been rejected by supporters of the SR/GR formalism.
As for space we note that it has a differential structure, in that different regions
are in relative motion.  This is caused by the gravitational in-flow effect locally, and 
as well by the growth of the universe.

\subsection{  In-Flow Turbulence and Gravitational Waves \label{subsection:inflowturbulence}}

\begin{figure}[ht]
\hspace{15mm}\includegraphics[scale=1.5]{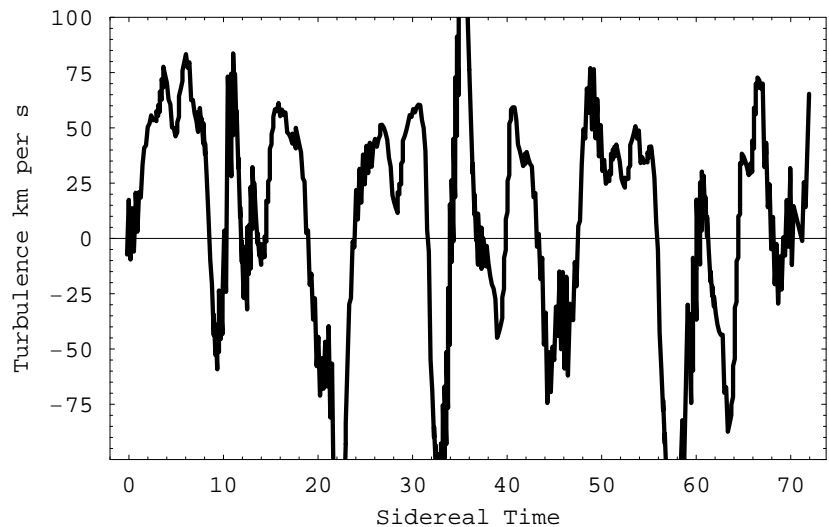}
\caption{\small{ Speed fluctuations  determined from Fig.\ref{fig:DeWittetimes}
by subtracting a least squares best fit of the forms  shown in Fig.\ref{fig:DeWitteTheory}b.
A 1ns variation in travel time corresponds approximately to a speed variation of 27km/s.  The
larger speed fluctuations  actually arise from a fluctuation in the cross-over time, that is, a
fluctuation in the direction of the velocity.  This plot implies that the velocity flow-field
is turbulent.  The scale of this turbulence is comparable to that evident in the Miller data, as
shown in Fig.\ref{fig:MillerData}  and Fig.\ref{fig:SeptPlot}. }  
\label{fig:Turbulence}}\end{figure}

The velocity  flow-field equation \cite{RCPP2003} is
expected to have solutions possessing turbulence, that is, random fluctuations in both the
magnitude and direction of the gravitational in-flow component of the velocity flow-field.   
Indeed all the Michelson interferometer experiments showed evidence of such turbulence. The first
clear evidence was from the Miller experiment, as shown in Fig.\ref{fig:MillerData}  and
Fig.\ref{fig:SeptPlot}.  Miller offered no explanation for these fluctuations  but in his
analysis of that data he did running time averages, as shown by the smoother curves in
Fig.\ref{fig:MillerData}.  Miller may have in fact have simply interpreted these fluctuations as
purely instrumental effects.  While some of these fluctuations may be partially caused by weather
related temperature and pressure  variations, the bulk of the fluctuations appear to be larger
than expected from that cause alone.   Even the original Michelson-Morley data in
Fig.\ref{fig:MMplots} shows variations in  the velocity field and supports this
interpretation.    However it is significant that the non-interferometer DeWitte data also shows
evidence of turbulence in both the magnitude and  direction of the velocity flow field, as shown
in Fig.\ref{fig:Turbulence}.  Just as the DeWitte data agrees
with the Miller data for speeds and directions the magnitude fluctuations, shown in
Fig.\ref{fig:Turbulence}, are very similar in absolute magnitude to, for example, the speed 
turbulence shown in Fig.\ref{fig:SeptPlot}.  

It therefore  becomes clear that there is
strong evidence for these fluctuations being evidence of  physical turbulence in the flow
field.  The magnitude of this turbulence appears to be somewhat larger than that which would be
caused by the in-flow of quantum foam towards the Sun, and indeed following on from
Sect.\ref{subsection:galacticinflow}  some of this turbulence may be associated with galactic
in-flow into the Milky Way.  This in-flow turbulence is a form of gravitational wave and the
ability of gas-mode Michelson interferometers to detect absolute motion means that experimental
evidence of such a wave phenomena has been available for a considerable period of time, but
suppressed along with the detection of absolute motion itself.   Of course flow equations  do not exhibit those
gravitational waves of the form that have been predicted to exist based on the Einstein equations, and which are
supposed to propagate at the speed of light.  All this means that gravitational wave phenomena is very easy to detect
and amounts to new physics that can be studied in much detail, particularly using the new 1st-order
interferometer discussed in \cite{RCPP2003}.

\subsection{ Vacuum Michelson Interferometers\label{subsection:vacuummichelson}}

Over the years vacuum-mode Michelson interferometer experiments have become increasing popular, although the motivation
for such experiments appears to be increasingly unclear. The first vacuum interferometer experiment was by Joos
\cite{vacuum} in 1930 and gave   $v_M<1$km/s.  This result is  consistent with a null effect as  predicted by both the
quantum-foam physics and  the Einstein physics.  Only Newtonian physics is disproved by such experiments.  These   vacuum
interferometer experiments do give null results, with increasing confidence level, as for   example in
Refs.\cite{KT,BH,Muller,NewVacuum}, but they only check that the  Lorentz contraction effect completely cancels the
geometrical path-length effect in vacuum experiments, and this is common to both theories. So they are unable to
distinguish the new physics from the Einstein physics.   Nevertheless  recent works \cite{Muller,NewVacuum} continue to
claim that the experiment had been motivated by the desire to look for evidence of absolute motion, despite effects of 
absolute motion  having been discovered as long ago as 1887. The `null results' are always reported as proof of the
Einstein formalism.  Of course all the vacuum experiments can do is check the Lorentz contraction effect, and this in
itself is valuable.  Unfortunately the analysis of the data from such experiments is always by means of the Robertson
\cite{Robertson} and  Mansouri and Sexl formalism \cite{MS}, which purports to be a generalisation of the Lorentz
transformation if there is a preferred frame.  However in \cite{RCPP2003} we have already
noted that absolute motion effects, that is the existence of a preferred frame, are consistent with the usual Lorentz
transformation, based as it is on the restricted Einstein measurement protocol.  A preferred frame is revealed by
gas-mode Michelson interferometer experiments, and then the refractive index of the gas plays a critical role in
interpreting the data.  The Robertson and  Mansouri-Sexl formalism contains no  contextual aspects such as a refractive
index effect and is thus totally inappropriate to the analysis of so called  `preferred frame' experiments.

It is a curious feature of the history of Michelson interferometer experiments that it went unnoticed that   the
results fell into two distinct classes, namely vacuum and gas-mode,  with recurring  non-null results from
gas-mode interferometers.   
\subsection{ Solid-State Michelson Interferometers\label{subsection:solidstate}}

The gas-mode Michelson interferometer has its sensitivity to absolute motion effects greatly reduced by the refractive
index effect, namely the $k^2=n^2-1$ factor in (\ref{eqn:QG0}), and for gases with $n$ only slightly greater than one this
factor has caused much confusion over the last 115 years.  So it would be expected that passing the light beams
through a transparent solid with  $n \approx 1.5$ rather than through a gas would greatly increase the sensitivity. 
Such an Michelson interferometer experiment was performed by Shamir and Fox \cite{ShamirFox} in Haifa in 1969. This
interferometer used light from a He-Ne laser and used perspex rods with  $L=0.26$m. The experiment was
interpreted  in terms of  the   supposed Fresnel drag effect, which has a drag coefficient given by
$b=1-1/n^2$. The light passing through the solid was supposed to be `dragged' along in the direction of 
motion of the solid with a velocity $\Delta {\bf V}=b{\bf v}$ additional to the usual $c/n$ speed. As well  
the Michelson geometrical path difference and the Lorentz contraction effects were incorporated into the
analysis.  The outcome was that no fringe shifts were seen on rotation of the interferometer, and Shamir and
Fox concluded that this negative result {\it``enhances the experimental basis of special relativity''}.

The Shamir-Fox experiment was unknown to us\footnote{This experiment was  performed by Professor Warren Lawrance, an
experimental physical chemist with considerable laser experience.} at Flinders university when in 2002  several meters of
optical fibre were used in a Michelson interferometer experiment which also used a He-Ne laser light source.  Again
because of the $n^2-1$ factor, and even ignoring the Fresnel drag effect, one would have expected large fringe shifts on
rotation of the interferometer, but none were observed.  As well in a repeat of the experiment single-mode optical fibres
were also used and again with no rotation effect seen.   So this experiment is consistent with the Shamir-Fox experiment. 
Re-doing the analysis by including the supposed   Fresnel drag effect, as Shamir and Fox did, makes no material
difference to the expected outcome.  In combination with the non-null results from the gas-mode interferometer
experiments along with the non-interferometer experiment of DeWitte it is clear that transparent solids behave differently
to a gas when undergoing absolute motion through the quantum foam.  Indeed this  in itself is a discovery of a new
phenomena.   

The most likely explanation is that the physical Fitzgerald-Lorentz contraction effect has a anisotropic effect on the
refractive index of the transparent solid, and this is such as to cause a cancellation of any differences in travel time
between the two arms  on rotation of the interferometer.  In this sense a transparent solid medium shares this
outcome with the vacuum itself.

\subsection{  Absolute Motion and Quantum Gravity\label{subsection:absolutmeqg}}
 
Absolute rotational motion had been recognised as a meaningful and obervable phenomena from the very beginning of
physics. Newton had used his rotating bucket experiment to illustrate the reality of absolute rotational motion,
and later Foucault  and Sagnac provided further experimental proof. But for absolute linear motion the history
would turm out to be completely different. It was generally thought that absolute linear motion was undetectable,
at least until Maxwell's electromagnetic theory appeared to require it. In perhaps the most bizarre sequence of
events in modern science it turns out that absolute linear motion has been apparent within experimental data for
over 100 years. It was missed in the first experiment designed to detect it and from then on for a variety of
sociological reasons it became a concept rejected  by physicists and banned from their journals despite
continuing  new experimental evidence. Those who pursued the scientific evidence were treated with scorn and
ridicule.  Even worse was  the impasse that this obstruction of the scientific process resulted in, namely the
halting of nearly all progress in furthering our understanding of the phenomena of gravity. For it is clear from
all the experiments that were capable of detecting absolute motion that there is present in that data evidence of
turbulence within the velocity field.  Both the in-flow itself and the  turbulence are manifestations at a
classical level of what is essentially  quantum gravity processes. 

Process Physics has given a unification of explanation and description of physical phenomena based upon the limitations of
formal syntactical systems which had nevertheless achieved a remarkable encapsulation of many phenomena, albeit in a
disjointed and confused manner, and with a dysfunctional ontology  attached for good measure.  As argued in early
sections space is a quantum system continually classicalised by on-going non-local collapse processes.  The emergent
phenomena is foundational to existence and experientialism.   Gravity in this system is caused by  differences in the
 rate of processing of the cellular information within the network which we experience as space, and
consequentially there is a differential flow of information which can be affected by the presence of matter or even by
space itself.  Of course the motion of matter including photons with respect to that spatial information content  is
detectable because it affects the geometrical and chronological attributes of that matter, and the experimental evidence
for this has been exhaustively discussed in this section.   What has become very clear is that the phenomena of gravity is
only understandable once we have this unification of the quantum phenomena of matter and the quantum phenomena of space
itself. In Process Physics the difference between matter and space is subtle. It comes down to the difference between
informational patterns that are topologically preserved and those information patterns that are not.   
One outcome of this unification is that as a consequence of having a quantum phenomena of space itself we obtain an
informational explanation for gravity, and which at a suitable level  has an emergent quantum description.  In this sense
we have an emergent quantum theory of gravity.  Of course no such quantum description of gravity  is derivable
from quantising Einsteinian gravity itself. This follows on two counts, one is that the Einstein gravity formalism 
fails on several levels, as discussed previously, and second that quantisation has no validity as a means of uncovering
deeper physics.   Most surprising of all is that   having uncovered the logical necessity for gravitational
phenomena it also appears that even the seemingly well-founded Newtonian account of gravity has major failings.  The
denial of this possibility has resulted in an unproductive search for dark matter.  Indeed like dark matter and
spacetime much of present day physics has all the hallmarks of another episode of Ptolemy's epicycles, namely concepts
that appear to be well founded but in the end turn out to be illusions, and ones that have acquired the status of dogma.

If the Michelson-Morley experiment had been properly analysed and the phenomena revealed by the data exposed, and this
would have required in 1887 that Newtonian physics be altered, then as well as the subsequent path of physics being very
different, physicists would almost certainly have discovered both the gravitational in-flow effect and associated  
 turbulence.  

It is clear then that observation and measurement of absolute motion leads directly to a
changed paradigm regarding the nature and manifestations of gravitational phenomena, and that the new 1st-order
interferometer described in \cite{RCPP2003}  will provide an extremely simple device to uncover
aspects of gravity previously denied by current physics.  There are two aspects of such an experimental program. One is
the characterisation of the turbulence and its linking to the new non-linear term in the velocity field theory. This is
a top down program. The second aspect is a bottom-up approach where the form of the velocity field theory, or its
modification, is derived from the deeper informational process physics.  This is essentially the quantum gravity
route.   The turbulence is of course essentially a gravitational wave phenomena and networks of 1st-order
interferometers will permit  spatial and time series analysis.  There are a number of other  gravitational anomalies
which may also  now be studied using such an interferometer network, and so much new physics can be expected to be
uncovered.

\section{  Conclusions\label{section:conclusions}}

We have shown here that six experiments so far have clearly revealed experimental evidence of absolute motion. As well
these are all consistent with respect to the direction and speed of that motion.  This clearly refutes the fundamental
postulates of the Einstein reinterpretation of the relativitsic effects that had been developed by Lorentz and others. 
Indeed these experiments are consistent with the Lorentzian relativity in which reality displays both absolute motion
effects and relativistic effects. As discussed in detail in \cite{RCPP2003} it is absolute motion that actually causes
these relativistic effects. As well these absolute motion experiments have given experimental support for a new theory
of gravity.  These developments are discussed more extensively in \cite{RCPP2003}.

\section{ References\label{section:references}}

\end{document}